\documentclass[aps,prx,twocolumn,superscriptaddress,showpacs,floatfix,longbibliography]{revtex4-2}
\usepackage{amsmath}
\usepackage{amssymb}
\usepackage{graphics}
\usepackage{graphicx}
\usepackage{float}
\usepackage{subfigure}
\usepackage{color}
\usepackage{hyperref}

\usepackage{bm}

\begin{document}
\title{Worldline deconfinement and emergent long-range interaction in the entanglement Hamiltonian and in the entanglement spectrum}

\author{Zenan Liu}
\affiliation{Department of Physics, School of Science and Research Center for Industries of the Future, Westlake University, Hangzhou 310030,  China}
\affiliation{Institute of Natural Sciences, Westlake Institute for Advanced Study, Hangzhou 310024, China}

\author{Zhe Wang}
\affiliation{Department of Physics, School of Science and Research Center for Industries of the Future, Westlake University, Hangzhou 310030,  China}
\affiliation{Institute of Natural Sciences, Westlake Institute for Advanced Study, Hangzhou 310024, China}

\author{Dao-Xin Yao}
\email{yaodaox@mail.sysu.edu.cn}
\affiliation{State Key Laboratory of Optoelectronic Materials and Technologies, Guangdong Provincial Key Laboratory of Magnetoelectric Physics and Devices, Institute of Neutron Science and Technology, School of Physics, Sun Yat-Sen University, Guangzhou 510275, China}

\author{Zheng Yan}
\email{zhengyan@westlake.edu.cn}
\affiliation{Department of Physics, School of Science and Research Center for Industries of the Future, Westlake University, Hangzhou 310030,  China}
\affiliation{Institute of Natural Sciences, Westlake Institute for Advanced Study, Hangzhou 310024, China}

\begin{abstract}

The entanglement spectrum (ES) is a powerful tool for probing topological phases. While its behavior in gapped systems is well understood, its properties in gapless regimes remain unclear. In this work, we employ a quantum Monte Carlo method to study the ES of a two-dimensional square-octagon lattice Heisenberg model at quantum criticality and in the Néel phase. We find that the ES exhibits an $M$-shape magnon mode with a distinct sublinear dispersion, deviating from the conventional linear magnon. This behavior, similar to that of a one-dimensional long-range Heisenberg chain, reveals the emergence of relevant long-range interactions in the entanglement Hamiltonian. We demonstrate that the mechanism underlying short- and long-range interactions in the entanglement Hamiltonian can be interpreted as the confinement/deconfinement of worldlines in the path integral formulation. Our results reveal that gapless modes can fundamentally change the entanglement Hamiltonian and its spectrum, thereby offering insight into this general phenomenon.

\end{abstract}
\date{\today}
\maketitle

\section{Introduction}
Entanglement is a unique and important concept in many-body physics, which has become a powerful tool to understand the topology and categorical symmetries of condensed matter~\cite{Calabrese2008entangle,Fradkin2006entangle,Nussinov2009Sufficient,Nussinov2009symmetry,Casini2007Universal,Ji2019Noninvertible,ji2020categorical,kong2020algebraic,Xiao2021Categorical,Xiao2021Universal,Jia2021Scaling,Jia2022Measuring,Wang2022scaling}. Its application also extends to the conformal field theory (CFT) \cite{Holzhey1994Geometric,Calabrese2004Entanglement}, quantum field theory, and quantum information~\cite{vidal2003entanglement,Korepin2004universality,Kitaev2006Topological,Levin2006Detecting}.
Commonly, the entanglement entropy (EE) is widely used to probe various phases and phase transitions, which can also capture the universal properties and anomaly of systems, such as continuous symmetry breaking, ground state degeneracy and critical behaviors \cite{Hastings2010Measuring, Kallin2014Corner,Jia2021Scaling,Jia2022Measuring}. In the topological phases, EE can reveal the topological features and quantum dimensions of topological excitation via scaling behaviors \cite{Kitaev2006Topological,isakov2011topological,Zhang2011Topological,Jia2021Scaling}.

In addition, the entanglement spectrum (ES) was proposed to extract the universal feature and CFT information of the topological ordered phase, capturing more information than EE~\cite{Pollmann2010entangle,Fidkowski2010Entanglement,Yao2010Entanglement,Xiao2012General,Canovi2014Dynamics,Luitz2014Participation,Luitz2014Shannon,Luitz2014Universal,Chung2014Entanglement,Pichler2016Measurement,Cirac2011Entanglement,Stoj2020Entanglement,guo2021entanglement,Grover2013Entanglement,Assaad2014Entanglement,Assaad2015Stable,Parisen2018Entanglement,yu2022conformal}. Li and Haldane first proposed that the low-lying entanglement spectrum is similar to the edge energy spectrum on the physical boundary in the topological phase, dubbed as the Li-Haldane conjecture~\cite{Li2008entangle}. Their work demonstrated that the general $\nu=5/2$ quantum fractional Hall states have the same low-lying ES with topological structure. Then, it was theoretically proved that there is a general correspondence between the entanglement spectrum of (2+1)$d$ chiral gapped topological states and energy spectrum on (1+1)$d$ edges~\cite{Xiao2012General}. Following numerical results also show that this conjecture still holds in many magnetic systems including topological states~\cite{Poilblanc2010entanglement,zyan2021entanglement,zhu2019reconstructing,Cirac2011Entanglement,Lou2011Entanglement,liu2024demonstrating,mao2023sampling}.

For a system with gapped bulk and gapless edge, such as the symmetry-protected topological (SPT) phase, the entanglement Hamiltonian (EH) is generally believed to be short-range, corresponding to the virtual edge Hamiltonian, part of the famous Li-Haldane conjecture~\cite{Li2008entangle,Xiao2012General}.
Sometimes, although the EH contains the long-range interaction, these exponentially decaying long-range terms are irrelevant. In such cases, the ES still resembles the edge spectrum~\cite{PhysRevB.88.245137,dalmonte2022entanglement,giudici2018entanglement,dalmonte2018quantum,wu2023classical,song2023different}. Because the long-range interaction can be ignored, the EH retains physical properties similar to the virtual edge Hamiltonian. The conclusion also holds when both bulk and edge are gapped, supported by the numerical evidence and wormhole effect in the path integral of the reduced density matrix~\cite{liu2024demonstrating}. All the above conclusions are based on a bipartite entangled cut, in which the reduced system has both bulk and edge. When we split a ladder/bilayer system into two chains/layers by entangled cut, gapped system can even experience the emergence of a long-range interacting EH~\cite{PhysRevA.103.043321,Li2024Relevant,wang2025sudden}. Thus we restrict the discussion of this work in the first condition, that is, a bipartite entangled cut with both bulk and edge. 

A natural question is what will happen to the EH when the system is fully gapless. Field theory analysis suggests that the EH in binary Bose-Einstein condensates may exhibit relevant long-range interactions~\cite{PhysRevB.88.245137}. 
In physical systems described by the original Hamiltonian, extensive studies have shown that strong long-range interactions can fundamentally alter intrinsic physics, even violating the Mermin-Wagner theorem.  The Mermin-Wagner theorem states that in one-dimension (1D) and two-dimensional (2D) spatial dimensions, it is impossible for a continuous symmetry to be spontaneously broken at finite temperature in systems with sufficiently short-range interactions~\cite{Mermin1966Absence,mermin1967absence}, which does not apply to the discrete symmetries or long-range systems. Strong long-range interactions may change the dispersion power in continuous symmetry systems~\cite{Song2023Dynamical,Song2024Quantum,zhao2023finite} or open a gap of an original Goldstone mode~\cite{diessel2022generalized,RevModPhys.95.035002,Song2023Dynamical}. Therefore, the feature of a spectrum can serve as an indicator of whether long-range interactions are relevant. But it is still a challenging task for numerical methods to extract the large scale spectrum of the EH in (2+1)$d$ system.

In this work, we utilize a recently developed quantum Monte Carlo (QMC) method based on path integral sampling of the reduced density matrix~\cite{zyan2021entanglement,liu2024demonstrating} to reveal the importance of long-range interaction in the ES,  using a two-dimensional square-octagon lattice (SOL) Heisenberg model as an example. When the bulk is at the quantum critical point (QCP) and in the N\'eel phase, we observe that the ES exhibits a $M$-shape magnon with sublinear dispersion rather than linear dispersion. This evolution  of the ES is similar to the excitation spectrum in the one-dimensional $S=1/2$ long-range Heisenberg chain, which can be confirmed by the finite-size scaling of dispersion power in our QMC simulation. The presence of long-range interaction gives rise to a sublinear dispersion with power $s\approx 0.2$ in the N\'eel state for the ES, which clearly deviates from the linear dispersion expected in the conventional magnon spectrum. Moreover, we demonstrates that this behavior can be explained by the mechanism of worldline deconfinement created along the gapless mode in the system. 

\begin{figure}[t!]
\centering
\includegraphics[width=0.5\textwidth]{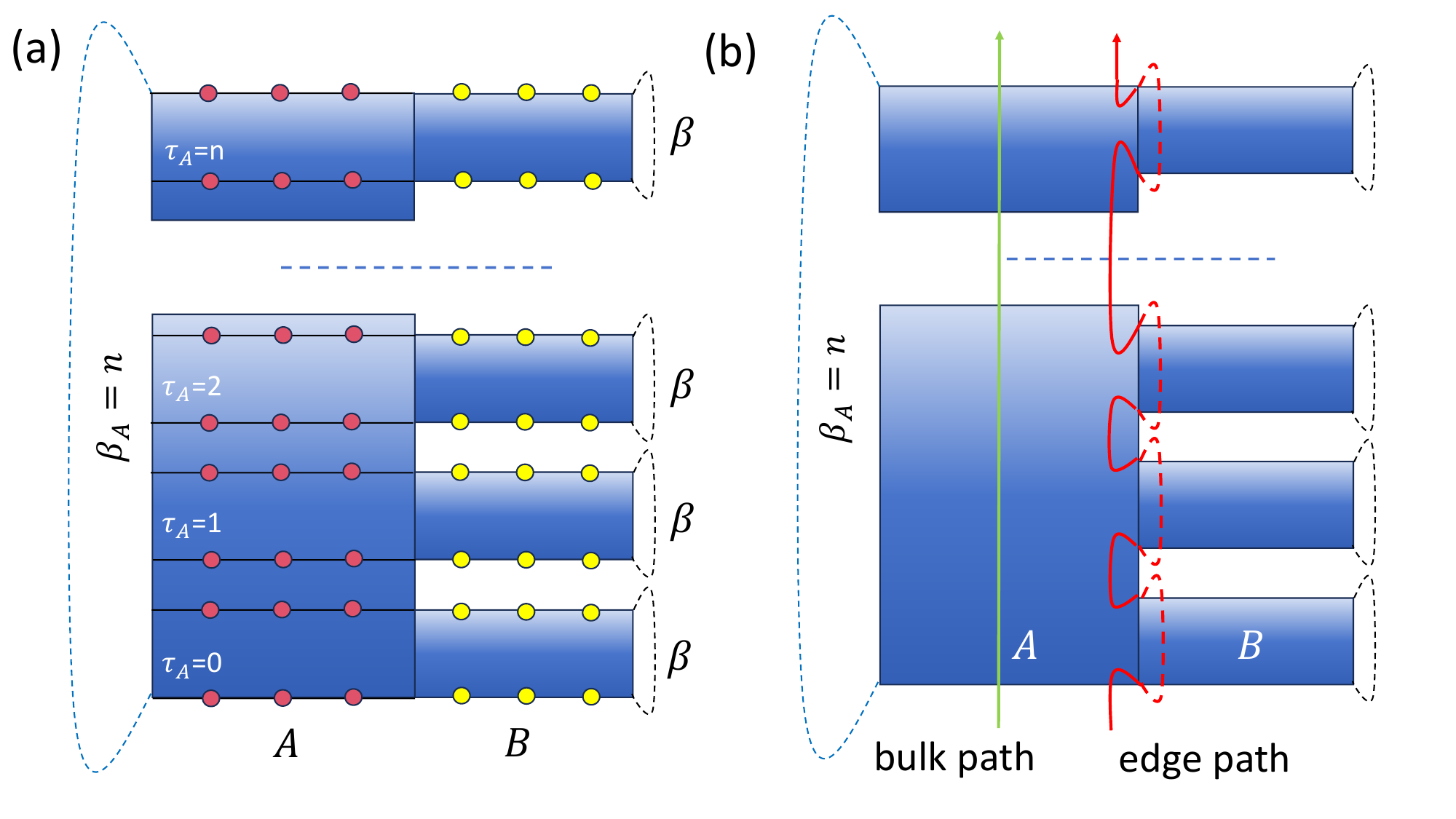}
\includegraphics[width=0.5\textwidth]{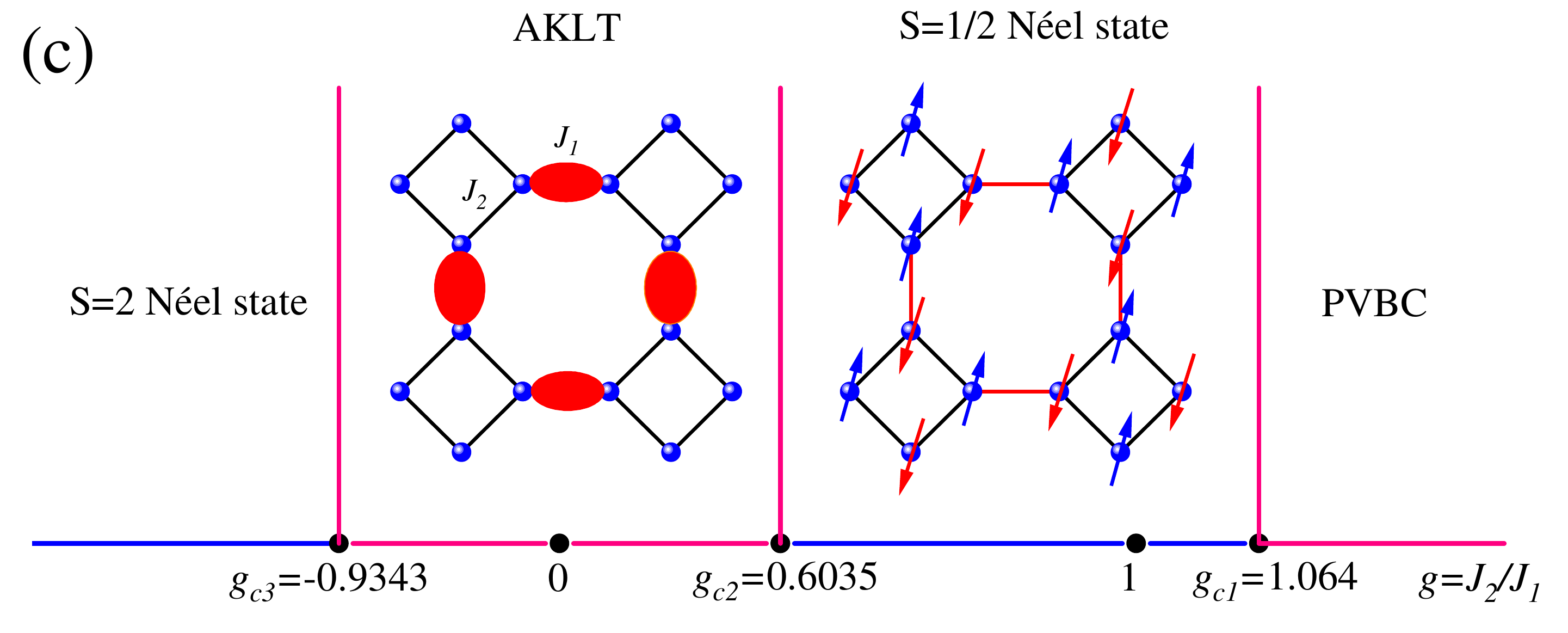}
\caption{(a) The replica manifold in the QMC simulation. The subsystem $A$ is entangled with the environment $\overline{A}$. Each replica connects with each other in $A$, while the environment $\overline{A}$ for each replica is independent. Here, $\beta_A = n$ denotes the imaginary-time length within $A$, and $\beta = 1/T$ corresponds to the inverse temperature (total imaginary-time length) of the full system. (b) The diagram for the wormhole mechanism. These worldlines which cross the replica via the bulk path will decay to zeros as $\beta \rightarrow \infty$ in principle. Meanwhile, the ones which cross the imaginary-time edge of $\overline{A}$ will arrive at the next replica without much cost. The dashed line represents a worldline jump to the next replica via the virtual path in the trace of environment, called the wormhole effect. (c) The phase diagram for the spin-1/2 Heisenberg model on the square-octagon lattice. We mainly focus on the AKLT phase and N\'eel phase.}
\label{fig-lattice}
\end{figure}

\section{Model and Method}
\subsection{model}
 We investigate the $S=1/2$ Heisenberg model on the square-octagon lattice via quantum Monte Carlo~\cite{Affleck1987Rigorous,zhang2017unconventional}.
\begin{equation}
\begin{split}
H = J_{1}\sum_{\langle ij\rangle}\mathbf{S}_{i}\cdot \mathbf{S}_{j}+J_{2}\sum_{\langle ij\rangle'}\mathbf{S}_{i}\cdot \mathbf{S}_{j}
\label{eq1}
\end{split}
\end{equation}
where $J_1$ is the inter-unit-cell interaction and $J_2$ is the intra-unit-cell interaction. We define $g=J_2/J_1$ with $J_1=1$. The model can host a rich phase diagram including the $S=2$ N\'eel phase, Affleck-Kennedy-Lieb-Tasaki (AKLT) phase, $S=1/2$ N\'eel phase and plaquatte valence bond crystal. The detailed phases and phase diagram can be found in Ref.~\cite{zhang2017unconventional}.
These phases are separated by three quantum critical points which all belong to the (2+1)$d$ $O(3)$ universality class, as shown in Fig.\ref{fig-lattice}(c). In the AKLT phase, four spin-1/2 on each plaquette couple to form an effective spin-2 degree of freedom. This effective spin-2 description makes the model amenable to quantum Monte Carlo simulations without the sign problem, serving as a low-energy effective theory of a spin-2 Heisenberg model on the square lattice. When the unit cell is defined as a plaquette including four spin-1/2, the system under open boundary condition exhibits the gapless edge mode--a hallmark of the AKLT phase. This behavior indicates that the presence of the AKLT phase intrinsically relies on the choice of unit cell. This boundary can be described by an effective $S=1/2$ Heisenberg chain which hosts gapless two-spinon continumm excitation~\cite{liu2022bulk} and exhibits quantum anomaly satisfying the Lieb-Schultz-Mattis theorem~\cite{lieb1961two,Oshikawa2000Commensurability,Hastings2004Lieb}. At the bulk quantum critical point, the boundary can induce exotic surface critical behaviors due to the coupling between edge mode and bulk critical mode~\cite{liu2024measuring}. This exotic QCP can be considered as gapless SPT or symmetry-enriched QCP~\cite{Scaffidi2017Gapless,Parker2018Topological,Verresen2021gapless,yu2022conformal,Yu2024Universal}, where the quantum anomaly still exists on the boundary at the QCP.

\subsection{Method}

 We use a recently proposed QMC method to extract the large-scale ES of two-dimensional SOL Heisenberg model. As we know, the reduced density matrix $\rho_A=\mathrm{Tr}_{\overline{A}}(|\psi\rangle\langle\psi|)=e^{-\mathcal{H}_{A}}$ can not be directly calculated via QMC simulation, though it is easy for exact diagonalization (ED) and density matrix renormalization
group (DMRG) to implement. Here, $A$ is the subsystem, $\bar{A}$ is the environment, and $\mathcal{H}_{A}$ is the corresponding entanglement Hamiltonian for the subsystem. Via the replica method~\cite{zyan2021entanglement}, we introduce an effective imaginary-time $n$ for the entanglement Hamiltonian and get the effective partition function of EH $\mathcal{H}_{A}$ as follows,
\begin{equation}
\begin{split}
\label{eq2}
\mathcal{Z}^{(n)}_{A}\propto \mathrm{Tr}[\rho^{n}_{A}]=\mathrm{Tr}[e^{-n \mathcal{H}_{A}}]
\end{split}
\end{equation}
where the replica number $n$ can be treated as the effective imaginary-time length $\tau_A$ for the EH (distinguished from the normal $\tau$, this $\tau_A$ is for the EH only), as shown in Fig.\ref{fig-lattice} (a). Actually, this method simulates the path integral of the reduce density matrix $\rho^n_A$. This approach allows
us to compute the entanglement spectrum for large-scale, high-dimensional quantum spin systems—a task that
is exceedingly difficult for established techniques like ED or DMRG.  It has successfully captured the entanglement spectra for 1D spin-1/2 Heisenberg ladder and two-dimensional bilayer Heisenberg model in large size systems~\cite{zyan2021entanglement}. 

The imaginary-time correlation of $\mathcal{H}_A$ can be obtained by measuring the $G(\tau_{A},\mathbf{q})=\langle O^{\dagger}_{-\mathbf{q}}(\tau_{A})O_{\mathbf{q}}(0)\rangle$, where $\tau_A$ is the imaginary time for the EH (the number of replica) and $\mathbf{q}$ is the momentum. As illustrated in Fig.\ref{fig-lattice}(a), $G(\tau_A)$ can be obtained by measuring the imaginary-time correlation function at integer point ($\tau_A=1,2,...,n$), as the imaginary-time difference between two nearest replica is fixed at $\delta \tau=1$. Then, we can extract the operator $O$ of spectral function $G(\omega,\mathbf{q})$ for the EH via stochastic analytic continuation (SAC) for the corresponding imaginary-time correlation ~\cite{Sandvik2016Constrained,Shao2023Progress,Zhou2021amplitude,yan2021topological}(SAC detail can be found in Appendix A). As a tool for studying dynamical properties of quantum many-body system, SAC has been extensively tested against synthetic data and compared with Bethe ansatz calculations, further confirming its reliability. Recently, it has been successfully applied to a range of  interesting systems, such as $S=1/2$ checkboard Heisenberg model~\cite{Xu2019}, $J-Q$ model~\cite{ma2018dynamical} and quantum spin liquid
~\cite{GYSun2018}, revealing insights into the physical feature of the excitation spectrum. Due to the translation symmetry and spin rotation symmetry, we focus on the boundary imaginary-time correlation function for the $s^z$ operator, $G(\tau_{A},q)=\frac{1}{L}\sum_{i,j}e^{-iq_{x}\cdot(x_{i}-x_{j})}\langle s^{z}_{i}(\tau)s^{z}_{j}(0)\rangle$, where $s^{z}_{i(j)}$ denotes spins on the entanglement boundary.

\begin{figure*}[htbp]
\centering
\includegraphics[width=0.9\textwidth]{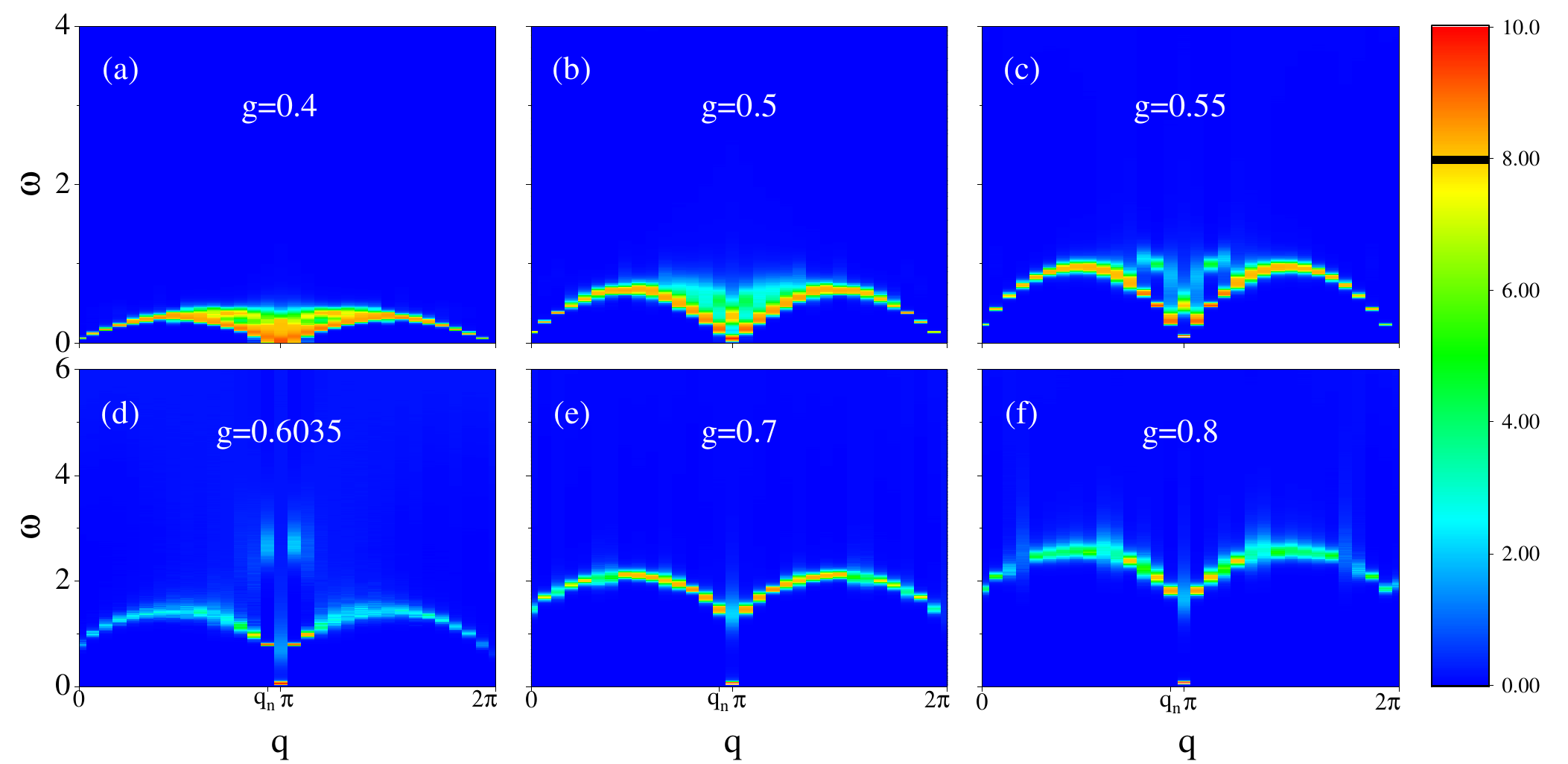}
\caption{Entanglement spectrum for the AKLT state and N\'eel state in the two-dimensional SOL Heisenberg model. $g$ is 0.4 (a), 0.5 (b), 0.55 (c), 0.6035 (d), 0.7 (e) and 0.8 (f). $q_n=15\pi/16$ is the closet discrete momentum to $q=\pi$ with $L=32$. For finite-size system $L$, $q_n=\pi-2\pi/L$ is the immediate momentum of $q=\pi$. For better presentation, we set the color bars to be logarithmic scale above 8.}
\label{Fig.s1}
\end{figure*}

\subsection{Wormhole picture}

 The new QMC method also gives a insight into the relationship between the energy spectrum and entanglement spectrum. In the replica manifold as shown in Figs.\ref{fig-lattice} (a) and (b), the trace of the environment actually connects the upper and lower edge of the replica as a periodic boundary condition in the imaginary-time direction. Thus it can provide a convenient way for woldlines near entangled edge to get to the upper edge from the lower edge very fast, dubbed as wormhole picture [here, 'wormhole' refers specifically to a path connecting different replicas in the extended spacetime of the replica trick, as shown in Fig.\ref{fig-lattice}(b)]. In the wormhole picture, worldlines can jump to the next replica via the virtual path
in the trace of environment (dashed line as shown in in Fig.\ref{fig-lattice}(b)). It makes worldlines near the entangled edge avoid crossing the bulk of the replica, as crossing would entail a huge cost, just as it does for worldlines deep in the bulk.

In the frame of the path integral, the cost is roughly proportional to the product of time-length and Hamiltonian gap on this path $\overline{\mathcal{L}} \times \Delta(\overline{\mathcal{L}})$, where $\overline{\mathcal{L}}$ is the average path length of imaginary time and $\Delta(\overline{\mathcal{L}})$ is the average gap for this path. The smaller cost will lead to larger path weight which can give greater contribution to the low-lying ES. Because the time-length in one replica is equal to $\beta$ and the time length around the entangled edge can be treated as a constant $1$ due to the wormhole effect, the ratio between the bulk path and edge path is roughly $\beta:1$.

When considering the path length and gap synthetically, the ratio of cost between bulk path and edge path can be estimated roughly as $\beta\Delta_b: \Delta_e$, where $\Delta_b$ is the bulk gap and $\Delta_e$ is the edge gap. $\beta$ should go to $\infty$ in the ground state which renders $\beta\Delta_b \gg \Delta_e$. Then the cost of the edge path is much lower than that of the  bulk path. The edge mode will have a larger weight and contribute more information to the low-lying ES. Therefore, the entangled edge mode in the EH is gapless as a virtual edge mode in the original Hamiltonian with open boundary condition; the Li-Haldane conjecture can be well explained via the wormhole picture~\cite{zyan2021entanglement}.

However, when the system is gapless, both $\Delta_b$ and $\Delta_e$ are zero, and the wormhole picture loses the effectiveness where the length of the imaginary-time path is irrelevant. We need to carefully calculate whether the worldlines are deconfined or confined in the path integral to detect the real costs of different paths.

\section{Entanglement spectrum}
We investigate the evolution of the entanglement spectrum when the system is from AKLT phase to $S=1/2$ N\'eel phase, as illustrated in Fig.\ref{Fig.s1}. The system size $L$ is 32 with $\beta_A=64$ and $\beta=50$. When $g=0.4<0.6035$, the system is in the deep AKLT phase. As shown in Fig.\ref{Fig.s1}(a), the ES is a gapless two-spinon continuum excitation, which is similar to the edge energy spectrum~\cite{liu2024demonstrating}. In this phase, the boundary can be considered as an effective 1D Luttinger liquid (here, Luttinger liquid phase is applied to describe the 1D effective boundary of 2D SOL model). When the $g$ becomes larger but still in the AKLT phase, such as $g=0.5$ and 0.55, the spinon continuum seems weaker, but it is hard to judge whether it has disappeared. As $g$ goes to the critical point $g_c= 0.6035$, the observed spectrum reveals a sharp dispersive mode with a likely gap at $q=15\pi/16$. This feature may be attributed to long-range interactions within the EH, as such interactions can suppress the dispersion power near $q=\pi$, leading to gap opening. However, confirming the presence of long-range interaction based solely on spectroscopic evidence remains challenging, which requires more finite-size scaling analysis to draw a definitive conclusion.

\begin{figure*}[htp]
\centering
\includegraphics[width=0.9\textwidth]{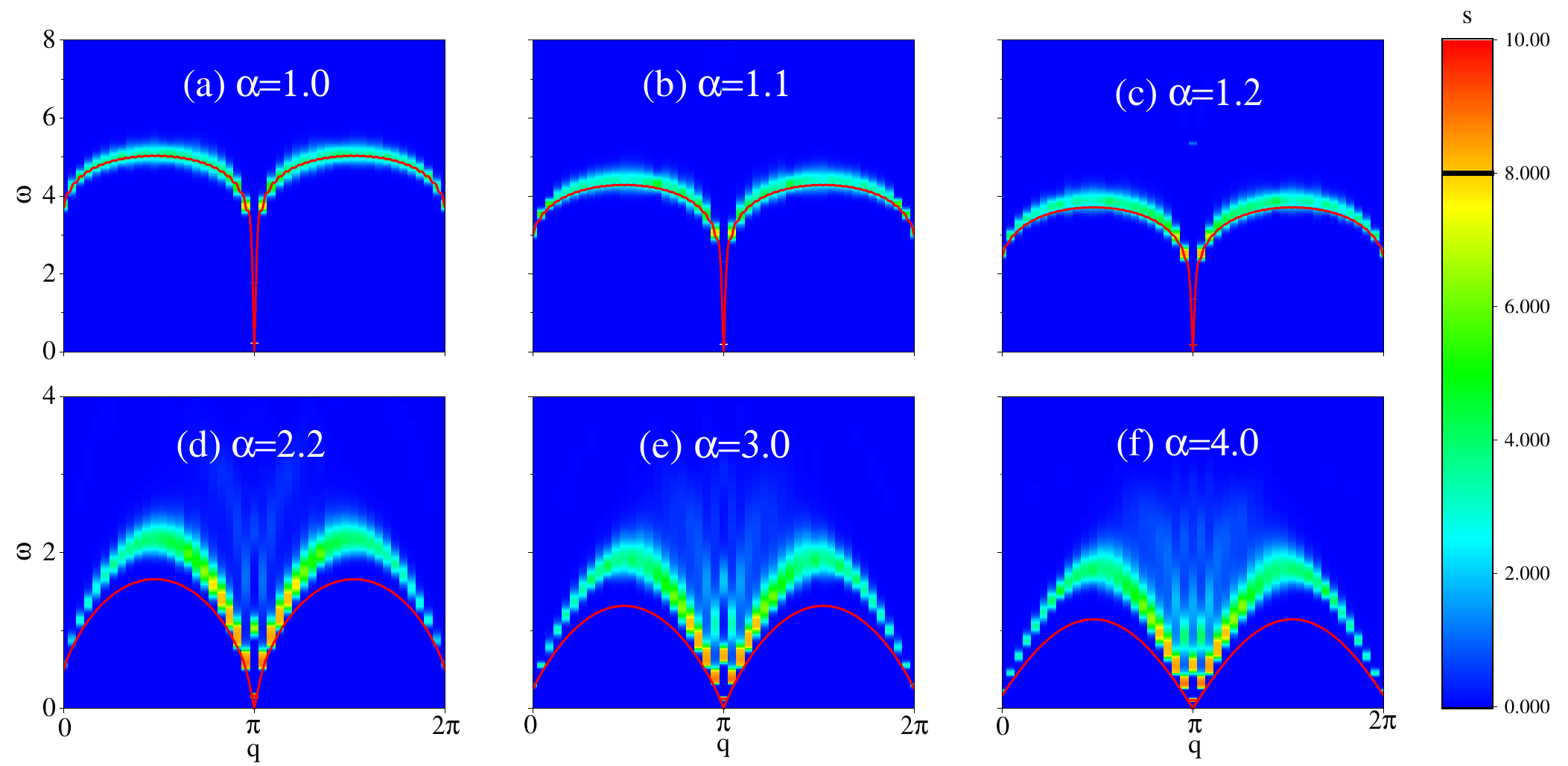}
\caption{Spin excitation spectra for the one-dimensional Heisenberg chain with long-range interaction from short-range area to long-range area for $L=48$ and $\beta=96$. The power law of long-range interaction $\alpha=1.0$(a), 1.1 (b), 1.2 (c), 2.2 (d), 3.0 (e) and 4.0 (f). The red lines denotes the linear spin wave theory results. For the Luttinger liquid ground state with $\alpha >2.23$, the spin wave results show deviations from the spectrum from QMC-SAC. For better presentation, we set the color bars to be logarithmic scale above 8.}
\label{Fig.s2}
\end{figure*}
When $g$ further increases, we observe that the two-spinon continuum obviously disappears. The ES becomes sharp magnon, which reflects that the bulk undergoes the quantum phase transition between AKLT phase and N\'eel phase. However, the dispersion near the $q=\pi$ becomes not linear as shown in Fig.\ref{Fig.s1}(d)-(f), which is not like the edge excitation in the N\'eel state with open boundary. In the previous work~\cite{liu2022bulk}, the real edge excitation is a gapless linear dispersion. 

Moreover, the gap near $q=\pi$ looks larger and larger as $g$ increases, which indicates that the ES may become discontinuous and opens a gap at $q=\pi-\frac{2\pi}{L}$ [Fig.\ref{Fig.s1}(d)-(f)]. As the bulk undergoes the phase transition, the length of the entanglement will increase. The short-range nature of the entanglement Hamiltonian will gradually change as bulk approaches the criticality. Therefore, the effective model for the subsystem should be no longer a short-ranged Hamiltonian, and the long-range interaction will be induced and become important. Since the physically original Hamiltonian is short-range for SOL Heisenberg model, the long-range interaction is emergent in the entanglement Hamiltonian as system enters into the gapless phase.

When $g>g_c$, the ES appears to open a gap at $q_n=\pi-2\pi/L$ ($15\pi/16$), which is primarily driven by the relevant long-range interaction in the entanglement Hamiltonian. Whether this gap remains finite in the thermodynamic limit remains to be checked by finite-size scaling.
This similar phenomenon can be also found in the two-dimensional Heisenberg model with long-range interaction on the square lattice~\cite{Song2023Dynamical,zhao2023finite}. These studies suggest that the entanglement Hamiltonian in the Néel phase can be effectively modeled as a Heisenberg chain with long-range interactions along the entanglement boundary. 

To further investigate the impact of long-range interaction on the spectrum, we consider a one-dimensional $S=1/2$ antiferromagnetic Heisenberg chain with power-law long-range interaction. This model allows us to systematically explore how long-range interactions influence the entanglement properties and spectral behaviors. This Hamiltonian is written as

\begin{equation}
\begin{split}
H = \sum_{i}\sum_{r\ge 1}J_{r}\mathbf{S}_{i}\cdot \mathbf{S}_{i+r}
\label{eq3}
\end{split}
\end{equation}
where $J_{r}=(-1)^{r-1}/r^{\alpha}$ without sign problem for QMC simulation. When $\alpha\rightarrow \infty$, this Hamiltonian becomes a short-ranged Heisenberg chain the ground state of which is Luttinger liquid (here, Luttinger liquid is applied to describe the ground state of the 1D system).  When $\alpha<3$ , the linear dispersion will break down due to the long-range interaction. As $\alpha$ gradually decreases, the long-range interaction will induce a long-range order. At $\alpha\approx 2.23$, the system undergoes a quantum phase transition between the N\'eel phase and Luttinger liquid phase~\cite{Laflorencie2005Critical,zhao2025Unconventional}. When $\alpha<2.23$, the long-range interaction becomes relevant and drives the system into the long-range N\'eel order~\cite{Yusuf2004spin,diessel2022generalized}. If $\alpha<1$, the spin excitation spectrum will become discontinous and open a gap near $q=\pi$, which enters the super long-range phase and falls in the fully connected Lieb-Mattis case~\cite{Lieb1962Ordering,zhao2025Unconventional}.

We extract the spin excitation spectrum of the one-dimensional Heisenberg chain with long-range interaction via QMC and SAC, as depicted in Fig.\ref{Fig.s2}. For large $\alpha$, the system is in a Luttinger liquid phase exhibiting a gapless two-spinon continuum. As $\alpha$ decreases, the growing long-range interaction suppresses and eventually confines the two-spinon continuum. Once the system enters the Néel phase, magnon excitations become dominant~\cite{yang2024dynamics}.

To benchmark the numerical results, we employ linear spin wave theory to obtain the low-energy dispersion of the long-range Heisenberg chain. Assuming magnetic order in the ground state, we apply the Holstein–Primakoff transformation at the linear level, expressing spin operators in terms of bosonic creation and annihilation operators, $S^{z}_{i}=S-a^{+}_{i}a_{i}, \quad S^{+}_{i}\approx\sqrt{2S}a_{i}, \quad S^{-}_{i}\approx\sqrt{2S}a^{+}_{i}$, and $S^{z}_{j}=b^{+}_{j}b_{j}-S, \quad S^{+}_{j}\approx\sqrt{2S}b^{+}_{j}, \quad S^{-}_{j}\approx\sqrt{2S}b_{j}$, where $a^{+}_{i}$ and $a_{i}$ are for up spins and $b^{+}_{j}$ and $b_{j}$ are for down spins. The resulting spin-wave Hamiltonian $H{\text{sw}}$ is diagonalized after Fourier transformation to yield the theoretical dispersion. As illustrated in Fig.~\ref{Fig.s2}, spin wave theory agrees well with QMC and SAC results inside the Néel phase ($\alpha < 2.23$, Fig.\ref{Fig.s2}(a-d)), but fails to capture the low-energy dispersion in the Luttinger liquid phase ($\alpha > 2.23$, Fig.\ref{Fig.s2}(e-f)).

A key observation emerges from comparing the ES in Fig.~\ref{Fig.s1} with the excitation spectrum of the long-range Heisenberg chain in Fig.~\ref{Fig.s2}. The evolution of the ES closely resembles that of the physical excitation spectrum as the interaction range is tuned. In both cases, the spectrum evolves from a broad two-spinon continuum in the short-range regime ($\alpha > \alpha_c$) to sharp, magnon-like modes in the long-range regime ($\alpha < \alpha_c$). This qualitative similarity between the ES and the spectrum of a long-range spin chain demonstrates that the entanglement Hamiltonian in the gapless phase incorporates emergent long-range interactions.

\begin{figure*}[htpb]
\centering
\includegraphics[width=0.90\textwidth]{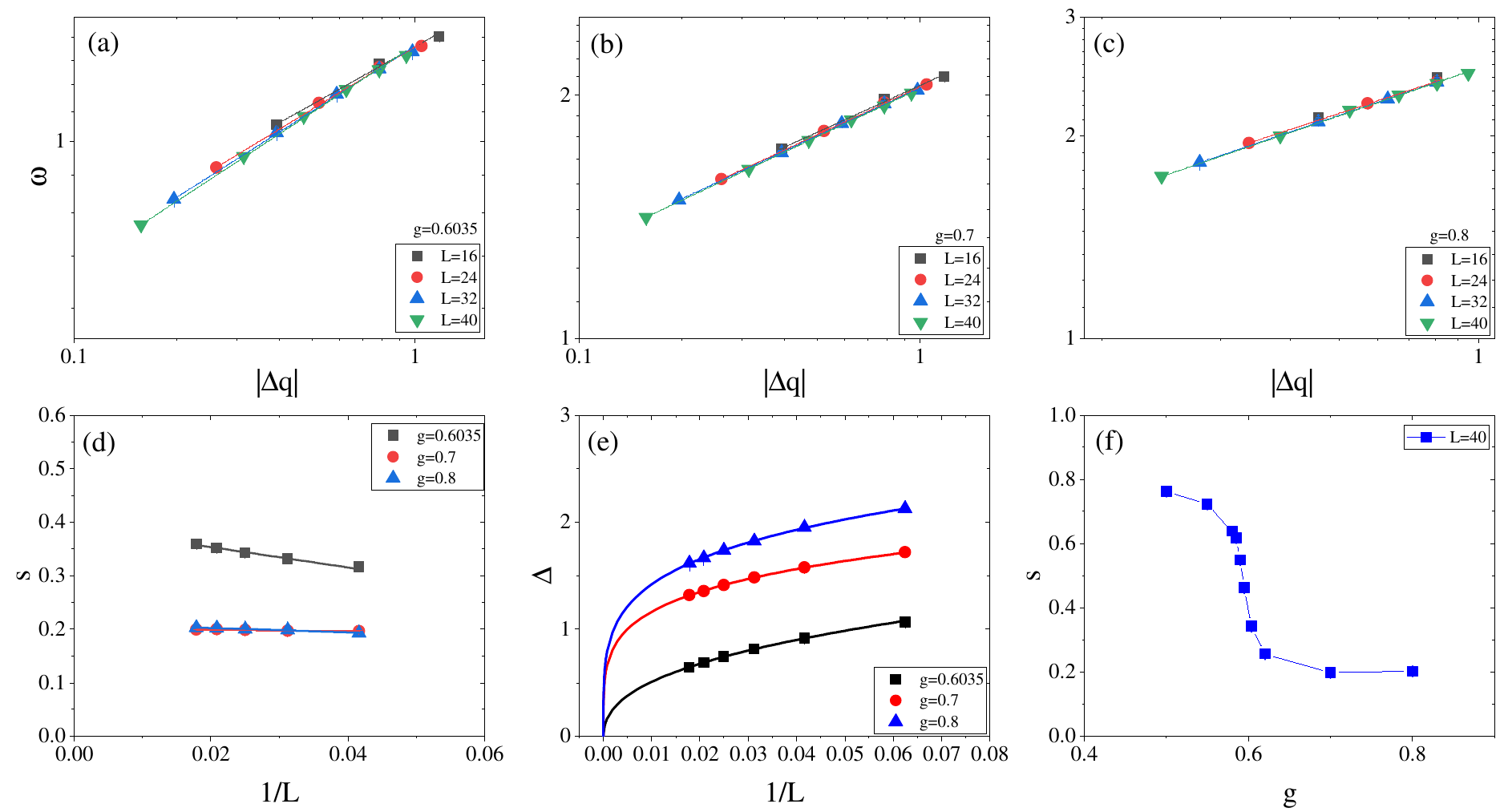}
\caption{Extracting the dispersion power $s$ from the dispersion curve with different $g$. (a)-(c): The dispersion curve obtained from QMC simulation with different $g$. (d) Finite-size scaling of the dispersion power $s$ fitting from (a) to (c), we fit the power $s$ by $\omega =a |\Delta q|^{s}$. (e) Finite size scaling of the entanglement gap $\Delta(q_n=\pi-2\pi/L)$ at $q_n=\pi-2\pi/L$. We use the function $\Delta=aL^s+\Delta(\infty)$ to fit the data with $s=0.3915$ ($g=0.6035$), 0.2016 ($g=0.7$) and 0.211($g=0.8$).  (f) Finite-size scaling of the dispersion power $s$  for different $g$ with system sizes $L=40$.}
\label{Fig.s3}
\end{figure*}

It is noteworthy that the dispersion near $q=\pi$ is not always linear and gapless. In the Néel phase, for instance, it becomes nearly discontinuous—a behavior also observed in the ES of the 2D SOL Heisenberg model (Figs. \ref{Fig.s1} and \ref{Fig.s2}). Spin wave theory predicts that near $q=\pi$, the dispersion follows a sublinear form:
\begin{align}
\omega(q) \propto |q-\pi|^{\frac{(\alpha-1)}{2}} (\alpha<3)
\label{eq4}
\end{align}
where exponent $s=(\alpha-1)/2$ from the spin-wave theory~\cite{Laflorencie2005Critical}. A similar expression applies near $q=0$. This sublinear behavior is also expected to appear in the ES of the N\'eel phase.  When $\alpha=1$, the dispersion power $s$ vanishes, leading to a gap spectrum near $q=\pi$ (Fig.\ref{Fig.s2}(a)). These findings suggest that the effective model governing the subsystem near the entanglement boundary can be described by a one-dimensional Heisenberg chain with relevant long-range interactions ($\alpha$ is small).

\begin{figure}[htp]
\centering
\includegraphics[width=0.5\textwidth]{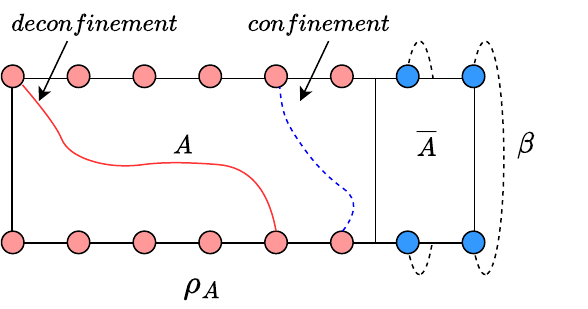}
\caption{The schematic diagram of worldline confinement and deconfinement in the reduced density matrix $\rho_A$. $A$ denotes the subsystem and $\overline{A}$ denotes the environment in the system. The blue dashed line denotes the worldline only connecting two spatially proximate spins, called worldline confinement.  The red solid line denotes the worldline connecting two spatially distant spins, called worldline deconfinement. }
\label{Fig.deconfine}
\end{figure}

\begin{figure*}[htpb]
\centering
\includegraphics[width=0.90\textwidth]{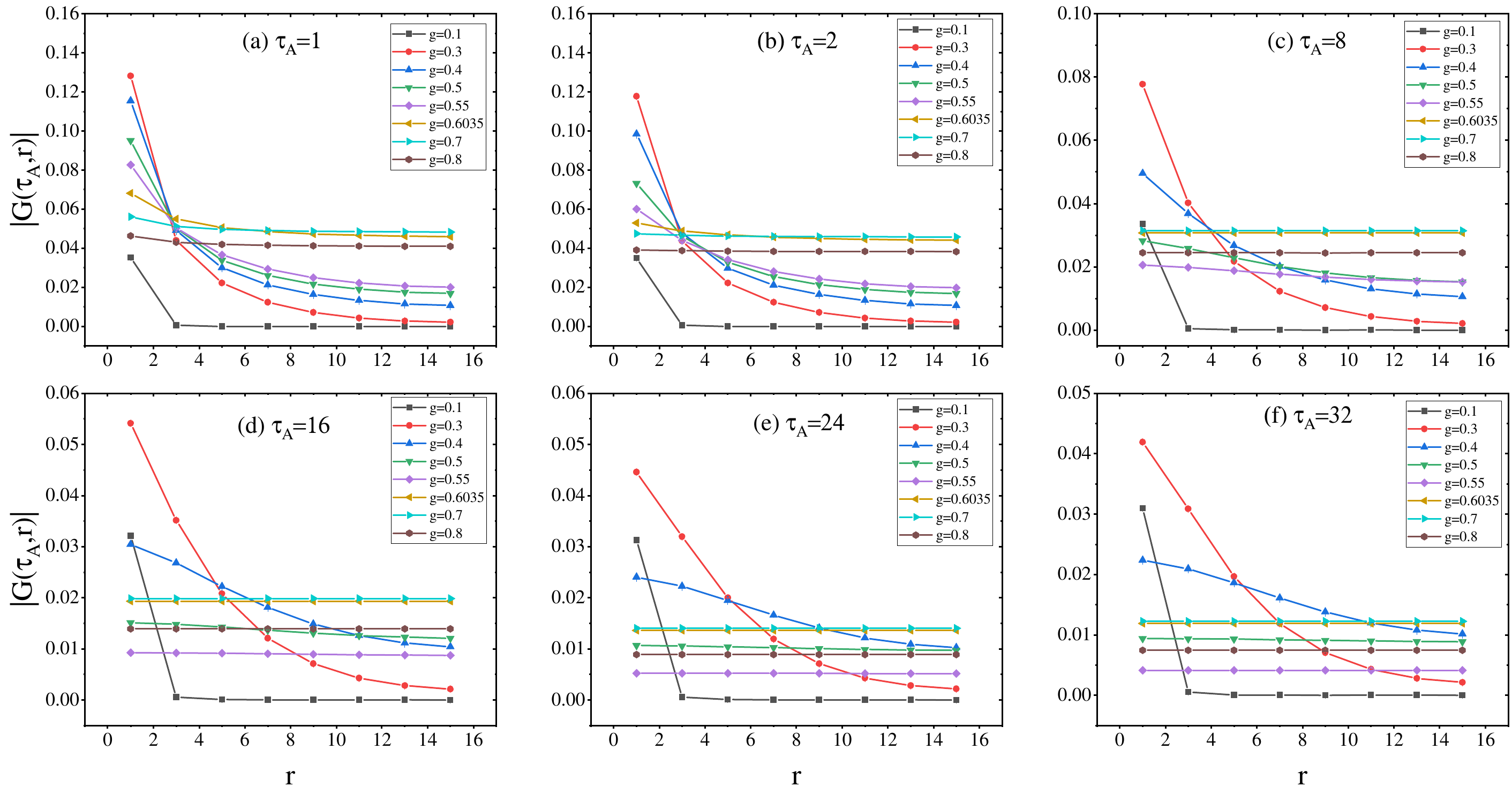}
\caption{Imaginary-time correlation function on the real space for different $g$ with fixed $\tau_A=1$ (a), 2 (b), 8 (c), 16 (d), 24 (e) and 32 (f).}
\label{Fig.s6}
\end{figure*}

\section{Dispersion power}
To determine whether the gap at $q_n=\pi-2\pi/L$ remains open in the thermodynamic limit, we perform a finite-size scaling analysis of both the dispersion exponent $s$ and the ES gap near $q=\pi$. We define $\Delta q=q-\pi$ and extract the exponent $s$ by fitting the dispersion to the form $\omega \propto |\Delta q|^s$. As shown in Figs.~\ref{Fig.s3}(a)-\ref{Fig.s3}(c), the dispersion near $q=\pi$ in the N\'eel phase follows Eq.~(\ref{eq4}) with $s<1$. At $g=0.7$ and 0.8, the extracted exponents $s\approx0.2$ indicate that long-range interactions strongly modify the dispersion relation. We further examine the system-size dependence of $s$ in Fig.~\ref{Fig.s3}(d), fitting the data to the form $s(L)=s(\infty)+bL^{-1}+O(L^{-2})$, where $s(\infty)$ is the dispersion power in the thermodynamic limit and $b$ is the fitting parameter. The extrapolated values are $s(\infty)=0.3915(2)$ for $g=0.6035$, $0.2016(7)$ for $g=0.7$, and $0.211(2)$ for $g=0.8$. Based on the dispersion relation for the 1D long-range Heisenberg chain, these results imply that the effective exponent $\alpha$ for the entanglement Hamiltonian satisfies $\alpha<2.23$, confirming that the long-range interaction term is relevant.

\begin{figure}[htpb]
\centering
\includegraphics[width=0.45\textwidth]{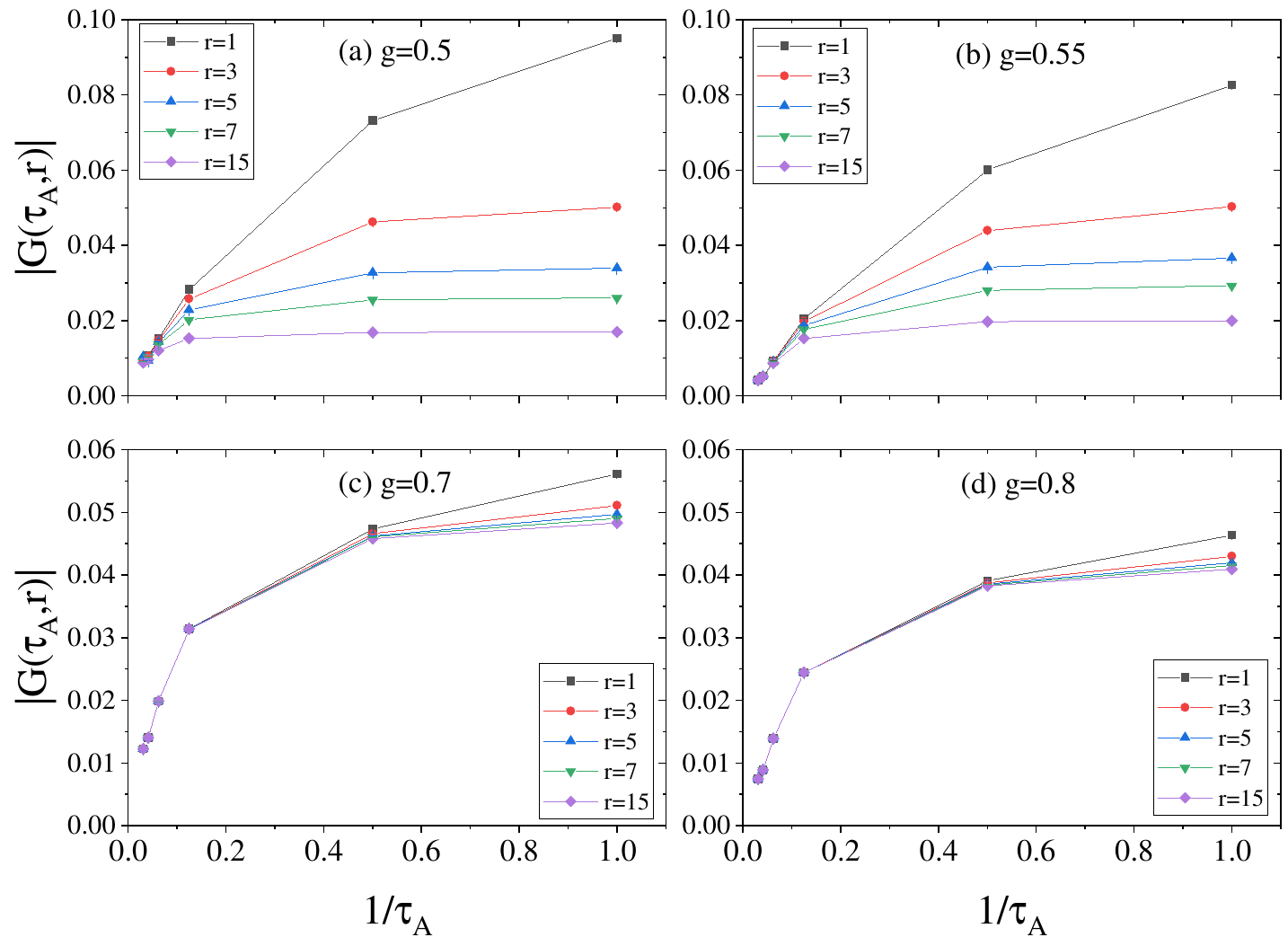}
\caption{Finite imaginary-time extrapolation of the imaginary-time correlation function on the real space for $g=0.5$ (a), 0.55 (b), 0.7 (c), and 0.8 (d) with fixed $r$.}
\label{Fig.s7}
\end{figure}

We further perform the finite-size analysis of the ES gap at $q_n=\pi-2\pi/L$. As shown in Fig.~\ref{Fig.s3}, the ES gap exhibits significant finite-size effects. If we fit the finite-size ES gap using polynomial functions (of orders 2–4), the extrapolated gap converges to a positive value as $L\rightarrow \infty$, suggesting the presence of a gap at $q_n=\pi-2\pi/L$. However, the scaling behavior should follow a power-law form $\Delta(L)=aL^s+\Delta(\infty)$, where $\Delta(\infty)$ is the gap in the thermodynamic limit and $s$ is the dispersion exponent. Since 
$s\ll1 $ in our case, the polynomial form is not appropriate.

When we instead use the power-law form with $s=0.3915$ to fit the data at $g=0.6035$, we obtain $\Delta(\infty)=-0.036(11)$, which is close to zero(Fig.~\ref{Fig.s3}(e)). For $g=0.7$ and $0.8$, using $s=0.216$ and 0.211 respectively, we find $\Delta(\infty)=-0.077(15)$ and $-0.081(22)$, both negative. This indicates that the gap converges slowly to zero, implying a gapless spectrum in the thermodynamic limit (Fig.~\ref{Fig.s3}(e)). Furthermore, assuming $\Delta(\infty)=0$ and fitting with $\Delta=aL^s$, we extract $s=0.4086(1)$ for $g=0.6035$, 0.2124(6) for $g=0.7$, and 0.221(1) for $g=0.8$. These values are consistent with the extrapolated results in Fig.~\ref{Fig.s3}(d), supporting the self-consistency of the power-law fitting.

We also extract the dispersion exponent $s$ for different $g$ values at $L=40$. As shown in Fig.~\ref{Fig.s3}(f), $s\approx 0.8$ for $g<0.6035$ in the AKLT phase, where long-range interactions are irrelevant. As the system approaches the quantum critical point and enters the Néel phase, $s$ decreases rapidly to about 0.2, indicating that long-range interactions become relevant near the entanglement boundary. This supports the conclusion that the entanglement Hamiltonian suddenly acquires long-range character when the ground state transitions from the gapped AKLT phase to the gapless N\'eel phase.

\section{Worldline deconfinement}
When the system is in the deep AKLT state, the EH can be effectively described by an $S=1/2$ short-ranged Heisenberg chain. Thus, the ES still hosts gapless two-spinon excitation which is analogous to a standard Heisenberg chain. Even when the boundary becomes gapped, the ES also looks like the edge energy spectrum, which can be understood by wormhole picture~\cite{liu2024demonstrating}. In the bulk gapped phase, the edge mode plays an important role in the ES due to its lower energy cost and greater weight in the replica manifold. Once the bulk enters a gapless phase, the gapless bulk excitations also provide a low-energy pathway for worldlines to propagate across replicas over long distances. This process is expected to induce non-local interactions in the entanglement Hamiltonian.

We propose a mechanism based on worldline behavior, termed worldline confinement/deconfinement, to explain the emergence of long-range interactions in the EH. In this picture, the range of interaction in the EH is closely related to the spatial extent that worldlines can span.

As illustrated in Fig.\ref{Fig.deconfine}, a worldline only connects two spatially nearby spins at different imaginary times, a behavior referred to as worldline confinement. If a worldline attempts to extend over a large distance, it must overcome a high energy barrier, reflecting the short-range nature of the EH in the AKLT phase. In contrast, at the quantum critical point or inside the Néel phase, a worldline can connect distant spins with relatively low energy cost, a situation termed worldline deconfinement. This behavior reveals the long-range feature of the entanglement
Hamiltonian in the gapless phase. In the gapped phase, the worldline is confined to regions near the entanglement boundary as a consequence of the wormhole effect. In contrast, for the gapless phase, the worldline is able to propagate freely even deep into the bulk, signaling the phenomenon of worldline deconfinement. 

Such a worldline connecting two spins can be interpreted as an imaginary-time off-diagonal operator of the form $S^{+}(\tau,r)S^{-}(0,0)$ and $S^{-}(\tau,r)S^{+}(0,0)$, which flips spins at different imaginary times and positions. The expectation of such operators corresponds to the off-diagonal imaginary-time correlation.  The relation between worldline and EH can be understood easily by the recently proposed method of sampling the reduced density matrix using quantum Monte Carlo~\cite{mao2023sampling}. In this approach, the off-diagonal elements of the reduced density matrix are proportional to the sampling frequency of worldlines that connect the basis states on the left and right sides.

Due to the $SU(2)$ symmetry for the EH, we can probe the worldline confinement and deconfinement by measuring the diagonal correlation $|G(\tau_A,r)|=|\langle S^z(\tau_A,r)S^z(0,0) \rangle|$ , since the diagonal operator satisfies $S^z(\tau,r)S^z(0,0) = \frac{1}{4}[S^{+}(\tau,r)S^{-}(0,0)+S^{-}(\tau,r)S^{+}(0,0)]$. When $\tau_A$ is small, the correlation function decays rapidly in the AKLT phase, whereas it saturates to a constant when the original system is at the critical point or in the Néel phase, as shown in Figs.\ref{Fig.s6}(a) and \ref{Fig.s6}(b). This behavior clearly demonstrates that worldlines are confined in the AKLT phase but become deconfined in the critical and Néel phases. However, the distinction between the AKLT and Néel phases becomes less clear at larger $\tau_A$. As illustrated in Figs.\ref{Fig.s6}(c)-\ref{Fig.s6}(f), for $g = 0.5$ and $0.55$, the imaginary-time correlation function gradually decays to a constant as $\tau_A$ increases, making it difficult to directly differentiate between worldline confinement in the AKLT phase and deconfinement in the Néel phase based solely on the correlation behaviors.

Notably, the entanglement spectrum obtained from QMC simulations reflects physical properties at large $\tau_A$ on the replica manifold. To clarify the asymptotic behavior, we perform a finite imaginary-time scaling analysis of the correlation function to determine whether worldline order develops in the AKLT phase. As shown in Fig.\ref{Fig.s7}, we observe that the correlation function slowly converges to zero in the AKLT phase (e.g., for $g = 0.5$ and $0.55$), while it saturates to a finite constant in the Néel phase. This result further confirms that worldlines remain confined in the AKLT phase but become deconfined in the critical and Néel phases, even in the limit $\tau_A \rightarrow \infty$.

Therefore, the emergent long-range interaction in the EH is both relevant and physically significant, which can be reflected directly in the confinement-deconfinement behavior of worldlines.

\section{CONCLUSION}
In summary, we investigate the ES of the two-dimensional SOL Heisenberg model. As the system evolves from AKLT state to N\'eel phase, the ES changes gradually from a gapless two-spinon continuum to sharp magnon excitation with a $M$-shape dispersion. This change in the ES is analogous to the spectral evolution in a long-range Heisenberg chain when transitioning from the short-range ($\alpha>\alpha_c$) to the long-range ($\alpha<\alpha_c$) regime. In the short-range Luttinger liquid phase ($\alpha>2.23$), the spectrum consists of a two-spinon continuum. When $\alpha$ decreases below $\alpha_c\approx 2.23$, the system enters the N\'eel phase, and the spectrum crosses over to sharp magnon modes. This remarkable agreement between the spectral crossover in the spin chain and the evolution of the ES in the SOL Heisenberg model strongly indicates that the entanglement Hamiltonian contains the relevant long-range interactions. Moreover, the dispersion does not keep linear and becomes sublinear, following a power law with exponent $s<1$, which is also analogous to the behavior observed in the one-dimensional long-range Heisenberg model. The ES gap at $q_n=\pi-2\pi/L$ is nearly open but ultimately closes in the thermodynamic limit, indicating that the long-range interaction in the entanglement Hamiltonian plays a crucial role both at the quantum critical point and in the N\'eel phase of the original systems. Our results demonstrate that the long-range interaction can significantly influence the entanglement Hamiltonian and spectrum, which should not be ignored in gapless phases. Furthermore, the emergence of such long-range interactions in the entanglement Hamiltonian can be interpreted through the mechanism of worldline deconfinement.

\section{Acknowledgements}
We thank F. Assaad and Z. Y. Meng for helpful discussions. D.X.Y. is supported by the National Key R\&D Program of China (Grant No. 2022YFA1402802), National Natural Science Foundation  of China (Grant Nos. 92565303, 92165204,12494591), Guang Dong Provincial Key Laboratory of Magnetoelectric Physics and Devices (Grant No. 2022B1212010008), Research Center for Magnetoelectric Physics of Guangdong Province (Grant No. 2024B0303390001), and Guangdong Provincial Quantum Science Strategic Initiative (Grant No. GDZX2401010). Z.L. is supported by the China Postdoctoral Science Foundation under Grant No.2024M762935 and NSFC Special Fund for Theoretical Physics under Grant No.12447119. Z.W. is supported by the China Postdoctoral Science Foundation under Grant No.~2024M752898. The work is supported by the the Scientific Research Project (Grant No. WU2025B011) and start-up funding of Westlake University. The authors thank the high-performance computing center of Westlake University and Beijng PARATERA Tech Co.,Ltd. for providing HPC resources.

\section{Data availability}
The data that support the findings of this article are openly available~\cite{liu2026zenodo}.

\newpage
\appendix

\section{Stochastic Analytic Continuation}
Stochastic analytic continuation is a powerful numerical tool to obtain the excitation spectrum $S(q,\omega)$ (dynamical structure factor). In order to capture $S(q,\omega)$, the imaginary-time correlation function should be measured by QMC simulation. The imaginary-time correlation function is given as

\begin{align}
G_{q}(\tau)=\langle \bm{S}_{-q}(\tau)\bm{S}_q(0)\rangle
\label{eq5}
\end{align}
When the Heisenberg spin is isotropic, $G_q(\tau)=3\langle S^z_{-q}(\tau)S^z_q(0)\rangle$, where $S^z_q$ is the Fourier form of the spin $S^z$.  For a set of imaginary-time points {$\tau_i$}, the statistic error of $G_q(\tau)$ with the same $q$ from QMC is correlated. Therefore, it is necessary to define the covariance matrix to express the full characterization. The covariance matrix is defined as follows

\begin{align}
C_{ij}=\frac{1}{N_b(N_b-1)}\sum^{N_b}_{\alpha=1}[G^{\alpha}(\tau_i)-\overline{G}(\tau_i)][(G^{\alpha}(\tau_j)-\overline{G}(\tau_j)]
\label{eq6}
\end{align}
Here $N_b$ is the number of QMC bins and $\overline{G}(\tau_i)$ is the statistical average of $G^{\alpha}(\tau_i)$.

The relationship between the imaginary-time correlation function and spectrum function in the boson systems is given as

\begin{align}
G(\tau)=\int_{0}^{\infty}\!d\omega\,S(q,\omega)\,K(\tau,\omega),  \nonumber \\
K(\tau,\omega)=\frac{e^{-\tau\omega}+e^{-(\beta-\tau)\omega}}{\pi\,(1+e^{-\beta\omega})},
\label{eq7}
\end{align}
where $K(\tau,\omega)$ is the boson kernel function. In the SAC process, the spectrum function is typically parametrized as the sum of a large number of $\delta$ functions, which can be classified into several distinct parametrization methods~\cite{Shao2023Progress}. This suitable spectrum function can be sampled in the Monte Carlo method using a likelihood function corresponding to the configuration weight,

\begin{align}
P(A)\propto \exp(-\frac{\chi^2}{2\Theta})
\end{align}
where $\chi^2$ is the goodness of fit and $\Theta$ is a fictitious temperature. The goodness of fit
is defined as
\begin{align}
\chi^2=\sum_{i,j}[G'(\tau_i)-\overline{G}(\tau_i)]C^{-1}_{i,j}[G'(\tau_j)-\overline{G}(\tau_j)],
\end{align}
where $G'(\tau_i)$ is calculated from the current spectrum function by Eq.(\ref{eq7}). The reliable results can be evaluated by $\chi^2$. The temperature adjustment scheme is implemented to find the suitable $\Theta$, as given in Ref. \cite{Shao2017nearly}. Our purpose is to adjust $\Theta$ to satisfy $\langle \chi^2\rangle \approx \chi^2_{min}+\sqrt{2\chi^2_{min}}$. The technical details can be found in Refs.\cite{Sandvik2016Constrained,Shao2023Progress,Shao2017nearly}. 

To obtain the entanglement spectrum with SAC, we should measure the imaginary-time correlation function along the entanglement edge in the generalized partition function $\mathcal{Z}^{(n)}_A\propto \mathrm{Tr}[\rho^{n}_{A}]=\mathrm{Tr}[e^{-n \mathcal{H}_{A}}]$. The momentum of spin operator $S_q$ we measure is given as

\begin{align}
S^z_q \equiv \frac{1}{\sqrt{N}}\sum_{i\in \partial A} e^{\,\mathrm{i}q\cdot r_i}\,S^z_i
\end{align}
where $\partial A$ denotes the entanglement boundary of the subsystem. Then, we can utilize this imaginary-time correlation functions near the entanglement boundary to obtain the entanglement spectrum.

For SAC, its primary error sources stem from two aspects: (i) the inherent statistical noise and finite precision of the input imaginary-time correlation functions; and (ii) the intrinsic limitations of the SAC algorithm itself. The quality and convergence of the spectrum also depend on the specific update strategy and sampling parameters used within the SAC procedure, which can introduce additional variations~\cite{Shao2023Progress}.

\section{Imaginary-time correlation function}

\begin{figure}[htp]
\centering
\includegraphics[width=0.45\textwidth]{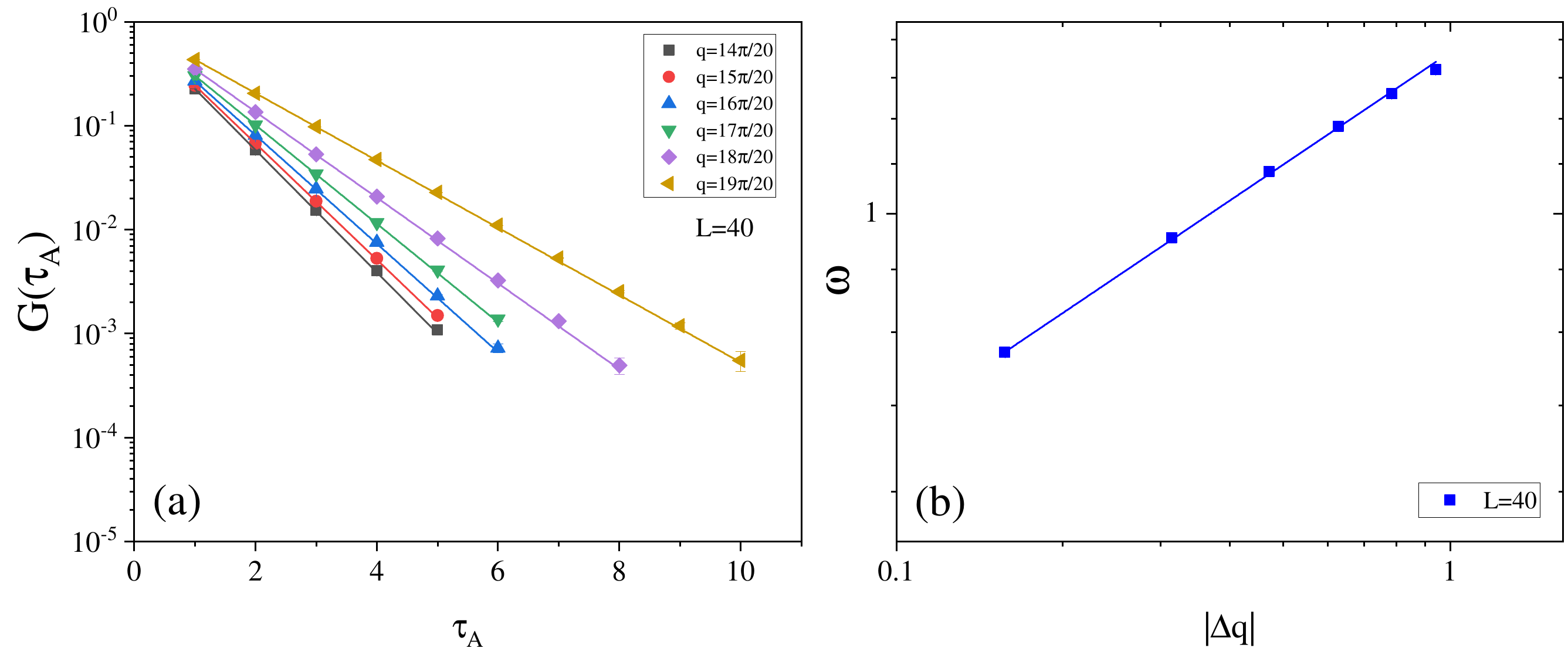}
\caption{(a) The imaginary-time correlation function $G(\tau_A)$ for the entanglement Hamiltonian in the $S=1/2$ square-octagon lattice Heisenberg model with $g=0.6035$ and $L=40$. The solid lines are fitting results. We obtain the gap of the entanglement Hamiltonian via the formula $G(\tau)=ae^{-\Delta\tau}$, where $\Delta$ refers to the gap. (b) The gap of the entanglement Hamiltonian near $q=\pi$, obtained from the data of (a). We fit the dispersion power near $q=\pi$ via $\omega =a |\Delta q|^s$.  }
\label{Fig.gtau}
\end{figure}

We obtain the gap of the EH at different $q$ by fitting the imaginary-ime correlation $G(\tau_A)$ in the entanglement Hamiltonian. As we know, the imaginary-time correlation function satisfies $G(\tau)=ae^{-\Delta\tau}$, where $\Delta$ refers to the gap. As shown in Fig.\ref{Fig.gtau}(a), $G(\tau_A)$ exhibits an exponential decay, reflecting the presence of a gap at different values of $q$.  By analyzing how the gap varies with $q$, we observe that its dispersion follows a power-law behavior, scaling as $\sim |\Delta q|^s$ ($s\approx 0.342$). This confirms that the dispersion exhibits a power-law behavior at small $q$.

\bibliography{AKLT}

@article{vidal2003entanglement,
  title = {Entanglement in Quantum Critical Phenomena},
  author = {Vidal, G. and Latorre, J. I. and Rico, E. and Kitaev, A.},
  journal = {Phys. Rev. Lett.},
  volume = {90},
  issue = {22},
  pages = {227902},
  numpages = {4},
  year = {2003},
  month = {Jun},
  publisher = {American Physical Society},
  doi = {10.1103/PhysRevLett.90.227902},
  url = {https://link.aps.org/doi/10.1103/PhysRevLett.90.227902}
}

@article{Korepin2004universality,
  title = {Universality of Entropy Scaling in One Dimensional Gapless Models},
  author = {Korepin, V. E.},
  journal = {Phys. Rev. Lett.},
  volume = {92},
  issue = {9},
  pages = {096402},
  numpages = {3},
  year = {2004},
  month = {Mar},
  publisher = {American Physical Society},
  doi = {10.1103/PhysRevLett.92.096402},
  url = {https://link.aps.org/doi/10.1103/PhysRevLett.92.096402}
}

@article{Wang2025Sudden,
  title = {Sudden change in the entanglement Hamiltonian: Phase diagram of an Ising entanglement Hamiltonian},
  author = {Wang, Zhe and Yang, Siyi and Mao, Bin-Bin and Cheng, Meng and Yan, Zheng},
  journal = {Phys. Rev. B},
  volume = {111},
  issue = {24},
  pages = {245126},
  numpages = {10},
  year = {2025},
  month = {Jun},
  publisher = {American Physical Society},
  doi = {10.1103/ywk9-n2ds},
  url = {https://link.aps.org/doi/10.1103/ywk9-n2ds}
}

@article{Levin2006Detecting,
	title = {Detecting Topological Order in a Ground State Wave Function},
	author = {Levin, Michael and Wen, Xiao-Gang},
	journal = {Phys. Rev. Lett.},
	volume = {96},
	issue = {11},
	pages = {110405},
	numpages = {4},
	year = {2006},
	month = {Mar},
	publisher = {American Physical Society},
	doi = {10.1103/PhysRevLett.96.110405},
	url = {https://link.aps.org/doi/10.1103/PhysRevLett.96.110405}
}

@article{Holzhey1994Geometric,
title = {Geometric and renormalized entropy in conformal field theory},
journal = {Nuclear Physics B},
volume = {424},
number = {3},
pages = {443-467},
year = {1994},
issn = {0550-3213},
doi = {https://doi.org/10.1016/0550-3213(94)90402-2},
url = {https://www.sciencedirect.com/science/article/pii/0550321394904022},
author = {Christoph Holzhey and Finn Larsen and Frank Wilczek},

}

@article{Calabrese2004Entanglement,
	doi = {10.1088/1742-5468/2004/06/p06002},
	url = {https://doi.org/10.1088/1742-5468/2004/06/p06002},
	year = 2004,
	month = {jun},
	publisher = {{IOP} Publishing},
	volume = {2004},
	number = {06},
	pages = {P06002},
	author = {Pasquale Calabrese and John Cardy},
	title = {Entanglement entropy and quantum field theory},
	journal = {Journal of Statistical Mechanics: Theory and Experiment},
}

@article{Calabrese2008entangle,
  title = {Entanglement spectrum in one-dimensional systems},
  author = {Calabrese, Pasquale and Lefevre, Alexandre},
  journal = {Phys. Rev. A},
  volume = {78},
  issue = {3},
  pages = {032329},
  numpages = {4},
  year = {2008},
  month = {Sep},
  publisher = {American Physical Society},
  doi = {10.1103/PhysRevA.78.032329},
  url = {https://link.aps.org/doi/10.1103/PhysRevA.78.032329}
}

@article{Nussinov2009Sufficient,
	author = "Nussinov, Zohar and Ortiz, Gerardo",
	title = "{Sufficient symmetry conditions for Topological Quantum Order}",
	doi = "10.1073/pnas.0803726105",
	journal = "Proc. Nat. Acad. Sci.",
	volume = "106",
	pages = "16944--16949",
	year = "2009"
}

@article{Nussinov2009symmetry,
	author = "Nussinov, Zohar and Ortiz, Gerardo",
	title = "{A symmetry principle for topological quantum order}",
	doi = "10.1016/j.aop.2008.11.002",
	journal = "Annals Phys.",
	volume = "324",
	pages = "977--1057",
	year = "2009"
}

@article{Casini2007Universal,
title = {Universal terms for the entanglement entropy in 2+1 dimensions},
journal = {Nuclear Physics B},
volume = {764},
number = {3},
pages = {183-201},
year = {2007},
issn = {0550-3213},
doi = {https://doi.org/10.1016/j.nuclphysb.2006.12.012},
url = {https://www.sciencedirect.com/science/article/pii/S0550321306010091},
author = {H. Casini and M. Huerta}
}

@article{Ji2019Noninvertible,
	title = {Noninvertible anomalies and mapping-class-group transformation of anomalous partition functions},
	author = {Ji, Wenjie and Wen, Xiao-Gang},
	journal = {Phys. Rev. Research},
	volume = {1},
	issue = {3},
	pages = {033054},
	numpages = {20},
	year = {2019},
	month = {Oct},
	publisher = {American Physical Society},
	doi = {10.1103/PhysRevResearch.1.033054},
}

@article{ji2020categorical,
	title = {Categorical symmetry and noninvertible anomaly in symmetry-breaking and topological phase transitions},
	author = {Ji, Wenjie and Wen, Xiao-Gang},
	journal = {Phys. Rev. Research},
	volume = {2},
	issue = {3},
	pages = {033417},
	numpages = {24},
	year = {2020},
	month = {Sep},
	publisher = {American Physical Society},
	doi = {10.1103/PhysRevResearch.2.033417},
	url = {https://link.aps.org/doi/10.1103/PhysRevResearch.2.033417}
}

@article{kong2020algebraic,
    title = {Algebraic higher symmetry and categorical symmetry: A holographic and entanglement view of symmetry},
    author = {Kong, Liang and Lan, Tian and Wen, Xiao-Gang and Zhang, Zhi-Hao and Zheng, Hao},
    journal = {Phys. Rev. Research},
    volume = {2},
    issue = {4},
    pages = {043086},
    numpages = {53},
    year = {2020},
    month = {Oct},
    publisher = {American Physical Society},
    doi = {10.1103/PhysRevResearch.2.043086},
    url = {https://link.aps.org/doi/10.1103/PhysRevResearch.2.043086}
}

@ARTICLE{Xiao2021Categorical,
    doi = {10.1088/1742-5468/ac08fe},
    url = {https://doi.org/10.1088/1742-5468/ac08fe},
    year = 2021,
    month = {jul},
    publisher = {{IOP} Publishing},
    volume = {2021},
    number = {7},
    pages = {073101},
    author = {Xiao-Chuan Wu and Wenjie Ji and Cenke Xu},
    title = {Categorical symmetries at criticality},
    journal = {Journal of Statistical Mechanics: Theory and Experiment}
}

@ARTICLE{Xiao2021Universal,
    title={Universal Features of Higher-Form Symmetries at Phase Transitions},
    author={Xiao-Chuan Wu and Chao-Ming Jian and Cenke Xu},
    journal={SciPost Phys.},
    volume={11},
    issue={2},
    pages={33},
    year={2021},
    publisher={SciPost},
    doi={10.21468/SciPostPhys.11.2.033},
    url={https://scipost.org/10.21468/SciPostPhys.11.2.033},
}

@ARTICLE{Jia2021Scaling,
  title = {Scaling of Entanglement Entropy at Deconfined Quantum Criticality},
author = {Zhao, Jiarui and Wang, Yan-Cheng and Yan, Zheng and Cheng, Meng and Meng, Zi Yang},
journal = {Phys. Rev. Lett.},
volume = {128},
issue = {1},
pages = {010601},
numpages = {6},
year = {2022},
month = {Jan},
publisher = {American Physical Society},
doi = {10.1103/PhysRevLett.128.010601},
url = {https://link.aps.org/doi/10.1103/PhysRevLett.128.010601}
}

@article{Jia2022Measuring,
  title={Measuring R{\'e}nyi entanglement entropy with high efficiency and precision in quantum Monte Carlo simulations},
  author={Zhao, Jiarui and Chen, Bin-Bin and Wang, Yan-Cheng and Yan, Zheng and Cheng, Meng and Meng, Zi Yang},
  journal={npj Quantum Materials},
  volume={7},
  number={1},
  pages={1--9},
  year={2022},
  publisher={Nature Publishing Group},
  doi = {10.1038/s41535-022-00476-0},
  url = {https://doi.org/10.1038/s41535-022-00476-0}
}

@Article{Wang2022scaling, 
	title={{Scaling of the disorder operator at deconfined quantum criticality}},
	author={Yan-Cheng Wang and Nvsen Ma and Meng Cheng and Zi Yang Meng},
	journal={SciPost Phys.},
	volume={13},
	pages={123},
	year={2022},
	publisher={SciPost},
	doi={10.21468/SciPostPhys.13.6.123},
	url={https://scipost.org/10.21468/SciPostPhys.13.6.123},
}

@article{Hastings2010Measuring,
	title = {Measuring Renyi Entanglement Entropy in Quantum Monte Carlo Simulations},
	author = {Hastings, Matthew B. and Gonz\'alez, Iv\'an and Kallin, Ann B. and Melko, Roger G.},
	journal = {Phys. Rev. Lett.},
	volume = {104},
	issue = {15},
	pages = {157201},
	numpages = {4},
	year = {2010},
	month = {Apr},
	publisher = {American Physical Society},
	doi = {10.1103/PhysRevLett.104.157201},
	url = {https://link.aps.org/doi/10.1103/PhysRevLett.104.157201}
}

@article{Kallin2014Corner,
doi = {10.1088/1742-5468/2014/06/P06009},
url = {https://dx.doi.org/10.1088/1742-5468/2014/06/P06009},
year = {2014},
month = {jun},
publisher = {IOP Publishing and SISSA},
volume = {2014},
number = {6},
pages = {P06009},
author = {A B Kallin and E M Stoudenmire and P Fendley and R R P Singh and R G Melko},
title = {Corner contribution to the entanglement entropy of an O(3) quantum critical point in 2 + 1 dimensions},
journal = {Journal of Statistical Mechanics: Theory and Experiment},
}

@article{isakov2011topological,
  title={Topological entanglement entropy of a Bose--Hubbard spin liquid},
  author={Isakov, Sergei V and Hastings, Matthew B and Melko, Roger G},
  journal={Nature Physics},
  volume={7},
  number={10},
  pages={772--775},
  year={2011},
  publisher={Nature Publishing Group UK London},
  doi = {10.1038/nphys2036},
  url = {https://doi.org/10.1038/nphys2036},
}

@article{Zhang2011Topological,
  title = {Topological entanglement entropy of ${\mathbb{Z}}_{2}$ spin liquids and lattice Laughlin states},
  author = {Zhang, Yi and Grover, Tarun and Vishwanath, Ashvin},
  journal = {Phys. Rev. B},
  volume = {84},
  issue = {7},
  pages = {075128},
  numpages = {7},
  year = {2011},
  month = {Aug},
  publisher = {American Physical Society},
  doi = {10.1103/PhysRevB.84.075128},
  url = {https://link.aps.org/doi/10.1103/PhysRevB.84.075128}
}

@article{Pollmann2010entangle,
  title = {Entanglement spectrum of a topological phase in one dimension},
  author = {Pollmann, Frank and Turner, Ari M. and Berg, Erez and Oshikawa, Masaki},
  journal = {Phys. Rev. B},
  volume = {81},
  issue = {6},
  pages = {064439},
  numpages = {10},
  year = {2010},
  month = {Feb},
  publisher = {American Physical Society},
  doi = {10.1103/PhysRevB.81.064439},
  url = {https://link.aps.org/doi/10.1103/PhysRevB.81.064439}
}

@article{Fidkowski2010Entanglement,
  title = {Entanglement Spectrum of Topological Insulators and Superconductors},
  author = {Fidkowski, Lukasz},
  journal = {Phys. Rev. Lett.},
  volume = {104},
  issue = {13},
  pages = {130502},
  numpages = {4},
  year = {2010},
  month = {Apr},
  publisher = {American Physical Society},
  doi = {10.1103/PhysRevLett.104.130502},
  url = {https://link.aps.org/doi/10.1103/PhysRevLett.104.130502}
}

@article{Yao2010Entanglement,
  title = {Entanglement Entropy and Entanglement Spectrum of the Kitaev Model},
  author = {Yao, Hong and Qi, Xiao-Liang},
  journal = {Phys. Rev. Lett.},
  volume = {105},
  issue = {8},
  pages = {080501},
  numpages = {4},
  year = {2010},
  month = {Aug},
  publisher = {American Physical Society},
  doi = {10.1103/PhysRevLett.105.080501},
  url = {https://link.aps.org/doi/10.1103/PhysRevLett.105.080501}
}

@article{Xiao2012General,
	title = {General Relationship between the Entanglement Spectrum and the Edge State Spectrum of Topological Quantum States},
	author = {Qi, Xiao-Liang and Katsura, Hosho and Ludwig, Andreas W. W.},
	journal = {Phys. Rev. Lett.},
	volume = {108},
	issue = {19},
	pages = {196402},
	numpages = {5},
	year = {2012},
	month = {May},
	publisher = {American Physical Society},
	doi = {10.1103/PhysRevLett.108.196402},
	url = {https://link.aps.org/doi/10.1103/PhysRevLett.108.196402}
}

@article{Canovi2014Dynamics,
  title = {Dynamics of entanglement entropy and entanglement spectrum crossing a quantum phase transition},
  author = {Canovi, Elena and Ercolessi, Elisa and Naldesi, Piero and Taddia, Luca and Vodola, Davide},
  journal = {Phys. Rev. B},
  volume = {89},
  issue = {10},
  pages = {104303},
  numpages = {12},
  year = {2014},
  month = {Mar},
  publisher = {American Physical Society},
  doi = {10.1103/PhysRevB.89.104303},
  url = {https://link.aps.org/doi/10.1103/PhysRevB.89.104303}
}

@article{Luitz2014Universal,
	title = {Universal Behavior beyond Multifractality in Quantum Many-Body Systems},
	author = {Luitz, David J. and Alet, Fabien and Laflorencie, Nicolas},
	journal = {Phys. Rev. Lett.},
	volume = {112},
	issue = {5},
	pages = {057203},
	numpages = {5},
	year = {2014},
	month = {Feb},
	publisher = {American Physical Society},
	doi = {10.1103/PhysRevLett.112.057203},
	url = {https://link.aps.org/doi/10.1103/PhysRevLett.112.057203}
}

@article{Luitz2014Shannon,
	title = {Shannon-R\'enyi entropies and participation spectra across three-dimensional $O(3)$ criticality},
	author = {Luitz, David J. and Alet, Fabien and Laflorencie, Nicolas},
	journal = {Phys. Rev. B},
	volume = {89},
	issue = {16},
	pages = {165106},
	numpages = {14},
	year = {2014},
	month = {Apr},
	publisher = {American Physical Society},
	doi = {10.1103/PhysRevB.89.165106},
	url = {https://link.aps.org/doi/10.1103/PhysRevB.89.165106}
}

@article{Luitz2014Participation,
	doi = {10.1088/1742-5468/2014/08/p08007},
	url = {https://doi.org/10.1088/1742-5468/2014/08/p08007},
	year = 2014,
	month = {aug},
	publisher = {{IOP} Publishing},
	volume = {2014},
	number = {8},
	pages = {P08007},
	author = {David J Luitz and Nicolas Laflorencie and Fabien Alet},
	title = {Participation spectroscopy and entanglement Hamiltonian of quantum spin models},
	journal = {Journal of Statistical Mechanics: Theory and Experiment}
}

@article{Chung2014Entanglement,
	title = {Entanglement spectroscopy using quantum Monte Carlo},
	author = {Chung, Chia-Min and Bonnes, Lars and Chen, Pochung and L\"auchli, Andreas M.},
	journal = {Phys. Rev. B},
	volume = {89},
	issue = {19},
	pages = {195147},
	numpages = {6},
	year = {2014},
	month = {May},
	publisher = {American Physical Society},
	doi = {10.1103/PhysRevB.89.195147},
	url = {https://link.aps.org/doi/10.1103/PhysRevB.89.195147}
}

@article{Pichler2016Measurement,
  title = {Measurement Protocol for the Entanglement Spectrum of Cold Atoms},
  author = {Pichler, Hannes and Zhu, Guanyu and Seif, Alireza and Zoller, Peter and Hafezi, Mohammad},
  journal = {Phys. Rev. X},
  volume = {6},
  issue = {4},
  pages = {041033},
  numpages = {12},
  year = {2016},
  month = {Nov},
  publisher = {American Physical Society},
  doi = {10.1103/PhysRevX.6.041033},
  url = {https://link.aps.org/doi/10.1103/PhysRevX.6.041033}
}

@article{Cirac2011Entanglement,
  title = {Entanglement spectrum and boundary theories with projected entangled-pair states},
  author = {Cirac, J. Ignacio and Poilblanc, Didier and Schuch, Norbert and Verstraete, Frank},
  journal = {Phys. Rev. B},
  volume = {83},
  issue = {24},
  pages = {245134},
  numpages = {12},
  year = {2011},
  month = {Jun},
  publisher = {American Physical Society},
  doi = {10.1103/PhysRevB.83.245134},
  url = {https://link.aps.org/doi/10.1103/PhysRevB.83.245134}
}

@article{Stoj2020Entanglement,
  title = {Entanglement-spectrum characterization of ground-state nonanalyticities in coupled excitation-phonon models},
  author = {Stojanovi\ifmmode \acute{c}\else \'{c}\fi{}, Vladimir M.},
  journal = {Phys. Rev. B},
  volume = {101},
  issue = {13},
  pages = {134301},
  numpages = {9},
  year = {2020},
  month = {Apr},
  publisher = {American Physical Society},
  doi = {10.1103/PhysRevB.101.134301},
  url = {https://link.aps.org/doi/10.1103/PhysRevB.101.134301}
}

@article{Grover2013Entanglement,
	title = {Entanglement of Interacting Fermions in Quantum Monte Carlo Calculations},
	author = {Grover, Tarun},
	journal = {Phys. Rev. Lett.},
	volume = {111},
	issue = {13},
	pages = {130402},
	numpages = {5},
	year = {2013},
	month = {Sep},
	publisher = {American Physical Society},
	doi = {10.1103/PhysRevLett.111.130402},
	url = {https://link.aps.org/doi/10.1103/PhysRevLett.111.130402}
}

@article{Assaad2014Entanglement,
  title = {Entanglement spectra of interacting fermions in quantum Monte Carlo simulations},
  author = {Assaad, Fakher F. and Lang, Thomas C. and Parisen Toldin, Francesco},
  journal = {Phys. Rev. B},
  volume = {89},
  issue = {12},
  pages = {125121},
  numpages = {7},
  year = {2014},
  month = {Mar},
  publisher = {American Physical Society},
  doi = {10.1103/PhysRevB.89.125121},
  url = {https://link.aps.org/doi/10.1103/PhysRevB.89.125121}
}

@article{Parisen2018Entanglement,
  title = {Entanglement Hamiltonian of Interacting Fermionic Models},
  author = {Parisen Toldin, Francesco and Assaad, Fakher F.},
  journal = {Phys. Rev. Lett.},
  volume = {121},
  issue = {20},
  pages = {200602},
  numpages = {6},
  year = {2018},
  month = {Nov},
  publisher = {American Physical Society},
  doi = {10.1103/PhysRevLett.121.200602},
  url = {https://link.aps.org/doi/10.1103/PhysRevLett.121.200602}
}

@article{Assaad2015Stable,
	title = {Stable quantum Monte Carlo simulations for entanglement spectra of interacting fermions},
	author = {Assaad, Fakher F.},
	journal = {Phys. Rev. B},
	volume = {91},
	issue = {12},
	pages = {125146},
	numpages = {7},
	year = {2015},
	month = {Mar},
	publisher = {American Physical Society},
	doi = {10.1103/PhysRevB.91.125146},
	url = {https://link.aps.org/doi/10.1103/PhysRevB.91.125146}
}

@article{yu2022conformal,
  title = {Conformal Boundary Conditions of Symmetry-Enriched Quantum Critical Spin Chains},
  author = {Yu, Xue-Jia and Huang, Rui-Zhen and Song, Hong-Hao and Xu, Limei and Ding, Chengxiang and Zhang, Long},
  journal = {Phys. Rev. Lett.},
  volume = {129},
  issue = {21},
  pages = {210601},
  numpages = {6},
  year = {2022},
  month = {Nov},
  publisher = {American Physical Society},
  doi = {10.1103/PhysRevLett.129.210601},
  url = {https://link.aps.org/doi/10.1103/PhysRevLett.129.210601}
}

@article{Lou2011Entanglement,
  title = {Entanglement spectra of the two-dimensional Affleck-Kennedy-Lieb-Tasaki model: Correspondence between the valence-bond-solid state and conformal field theory},
  author = {Lou, Jie and Tanaka, Shu and Katsura, Hosho and Kawashima, Naoki},
  journal = {Phys. Rev. B},
  volume = {84},
  issue = {24},
  pages = {245128},
  numpages = {9},
  year = {2011},
  month = {Dec},
  publisher = {American Physical Society},
  doi = {10.1103/PhysRevB.84.245128},
  url = {https://link.aps.org/doi/10.1103/PhysRevB.84.245128}
}

@article{Affleck1987Rigorous,
  title = {Rigorous results on valence-bond ground states in antiferromagnets},
  author = {Affleck, Ian and Kennedy, Tom and Lieb, Elliott H. and Tasaki, Hal},
  journal = {Phys. Rev. Lett.},
  volume = {59},
  issue = {7},
  pages = {799--802},
  numpages = {0},
  year = {1987},
  month = {Aug},
  publisher = {American Physical Society},
  doi = {10.1103/PhysRevLett.59.799},
  url = {https://link.aps.org/doi/10.1103/PhysRevLett.59.799}
}

@article{liu2022bulk,
  title = {Bulk and edge dynamics of a two-dimensional Affleck-Kennedy-Lieb-Tasaki model},
  author = {Liu, Zenan and Li, Jun and Huang, Rui-Zhen and Li, Jun and Yan, Zheng and Yao, Dao-Xin},
  journal = {Phys. Rev. B},
  volume = {105},
  issue = {1},
  pages = {014418},
  numpages = {7},
  year = {2022},
  month = {Jan},
  publisher = {American Physical Society},
  doi = {10.1103/PhysRevB.105.014418},
  url = {https://link.aps.org/doi/10.1103/PhysRevB.105.014418}
}

@article{Shao2017nearly,
  title = {Nearly Deconfined Spinon Excitations in the Square-Lattice Spin-$1/2$ Heisenberg Antiferromagnet},
  author = {Shao, Hui and Qin, Yan Qi and Capponi, Sylvain and Chesi, Stefano and Meng, Zi Yang and Sandvik, Anders W.},
  journal = {Phys. Rev. X},
  volume = {7},
  issue = {4},
  pages = {041072},
  numpages = {26},
  year = {2017},
  month = {Dec},
  publisher = {American Physical Society},
  doi = {10.1103/PhysRevX.7.041072},
  url = {https://link.aps.org/doi/10.1103/PhysRevX.7.041072}
}

@article{Sandvik2016Constrained ,
  title = {Constrained sampling method for analytic continuation},
  author = {Sandvik, Anders W.},
  journal = {Phys. Rev. E},
  volume = {94},
  issue = {6},
  pages = {063308},
  numpages = {5},
  year = {2016},
  month = {Dec},
  publisher = {American Physical Society},
  doi = {10.1103/PhysRevE.94.063308},
  url = {https://link.aps.org/doi/10.1103/PhysRevE.94.063308}
}

@article{wu2023classical,
  title = {Classical model emerges in quantum entanglement: Quantum Monte Carlo study for an Ising-Heisenberg bilayer},
  author = {Wu, Siying and Ran, Xiaoxue and Yin, Binbin and Li, Qi-Fang and Mao, Bin-Bin and Wang, Yan-Cheng and Yan, Zheng},
  journal = {Phys. Rev. B},
  volume = {107},
  issue = {15},
  pages = {155121},
  numpages = {8},
  year = {2023},
  month = {Apr},
  publisher = {American Physical Society},
  doi = {10.1103/PhysRevB.107.155121},
  url = {https://link.aps.org/doi/10.1103/PhysRevB.107.155121}
}

@article{Shao2023Progress,
title = {Progress on stochastic analytic continuation of quantum Monte Carlo data},
journal = {Physics Reports},
volume = {1003},
pages = {1-88},
year = {2023},
issn = {0370-1573},
doi = {https://doi.org/10.1016/j.physrep.2022.11.002},
url = {https://www.sciencedirect.com/science/article/pii/S0370157322003921},
author = {Hui Shao and Anders W. Sandvik},

}

@article{Kitaev2006Topological,
	title = {Topological Entanglement Entropy},
	author = {Kitaev, Alexei and Preskill, John},
	journal = {Phys. Rev. Lett.},
	volume = {96},
	issue = {11},
	pages = {110404},
	numpages = {4},
	year = {2006},
	month = {Mar},
	publisher = {American Physical Society},
	doi = {10.1103/PhysRevLett.96.110404},
	url = {https://link.aps.org/doi/10.1103/PhysRevLett.96.110404}
}

@article{zhao2023finite,
  title={Finite-temperature critical behaviors in 2D long-range quantum Heisenberg model},
  author={Zhao, Jiarui and Song, Menghan and Qi, Yang and Rong, Junchen and Meng, Zi Yang},
  journal={npj Quantum Materials},
  volume={8},
  number={1},
  pages={59},
  year={2023},
  publisher={Nature Publishing Group UK London},
  doi = {10.1038/s41535-023-00591-6}
}

@article{Song2024Quantum,
  title = {Quantum criticality and entanglement for the two-dimensional long-range Heisenberg bilayer},
  author = {Song, Menghan and Zhao, Jiarui and Qi, Yang and Rong, Junchen and Meng, Zi Yang},
  journal = {Phys. Rev. B},
  volume = {109},
  issue = {8},
  pages = {L081114},
  numpages = {8},
  year = {2024},
  month = {Feb},
  publisher = {American Physical Society},
  doi = {10.1103/PhysRevB.109.L081114},
  url = {https://link.aps.org/doi/10.1103/PhysRevB.109.L081114}
}

@article{dalmonte2018quantum,
  title={Quantum simulation and spectroscopy of entanglement Hamiltonians},
  author={Dalmonte, Marcello and Vermersch, Beno{\^\i}t and Zoller, Peter},
  journal={Nature Physics},
  volume={14},
  number={8},
  pages={827--831},
  year={2018},
  publisher={Nature Publishing Group UK London}
}

@article{giudici2018entanglement,
  title={Entanglement Hamiltonians of lattice models via the Bisognano-Wichmann theorem},
  author={Giudici, G and Mendes-Santos, Tiago and Calabrese, P and Dalmonte, Marcello},
  journal={Physical Review B},
  volume={98},
  number={13},
  pages={134403},
  year={2018},
  publisher={APS}
}

@article{liu2024measuring,
  title = {Measuring the Boundary Gapless State and Criticality via Disorder Operator},
  author = {Liu, Zenan and Huang, Rui-Zhen and Wang, Yan-Cheng and Yan, Zheng and Yao, Dao-Xin},
  journal = {Phys. Rev. Lett.},
  volume = {132},
  issue = {20},
  pages = {206502},
  numpages = {7},
  year = {2024},
  month = {May},
  publisher = {American Physical Society},
  doi = {10.1103/PhysRevLett.132.206502},
  url = {https://link.aps.org/doi/10.1103/PhysRevLett.132.206502}
}

@article{liu2024demonstrating,
  title = {Demonstrating the wormhole mechanism of the entanglement spectrum via a perturbed boundary},
  author = {Liu, Zenan and Huang, Rui-Zhen and Yan, Zheng and Yao, Dao-Xin},
  journal = {Phys. Rev. B},
  volume = {109},
  issue = {9},
  pages = {094416},
  numpages = {10},
  year = {2024},
  month = {Mar},
  publisher = {American Physical Society},
  doi = {10.1103/PhysRevB.109.094416},
  url = {https://link.aps.org/doi/10.1103/PhysRevB.109.094416}
}

@article{zhang2017unconventional,
  title={Unconventional surface critical behavior induced by a quantum phase transition from the two-dimensional affleck-kennedy-lieb-tasaki phase to a n{\'e}el-ordered phase},
  author={Zhang, Long and Wang, Fa},
  journal={Physical Review Letters},
  volume={118},
  number={8},
  pages={087201},
  year={2017},
  publisher={APS}
}

@article{RevModPhys.95.035002,
  title = {Long-range interacting quantum systems},
  author = {Defenu, Nicol\`o and Donner, Tobias and Macr\`{\i}, Tommaso and Pagano, Guido and Ruffo, Stefano and Trombettoni, Andrea},
  journal = {Rev. Mod. Phys.},
  volume = {95},
  issue = {3},
  pages = {035002},
  numpages = {70},
  year = {2023},
  month = {Aug},
  publisher = {American Physical Society},
  doi = {10.1103/RevModPhys.95.035002},
  url = {https://link.aps.org/doi/10.1103/RevModPhys.95.035002}
}

@article{PhysRevB.88.245137,
  title = {Entanglement spectra between coupled Tomonaga-Luttinger liquids: Applications to ladder systems and topological phases},
  author = {Lundgren, Rex and Fuji, Yohei and Furukawa, Shunsuke and Oshikawa, Masaki},
  journal = {Phys. Rev. B},
  volume = {88},
  issue = {24},
  pages = {245137},
  numpages = {14},
  year = {2013},
  month = {Dec},
  publisher = {American Physical Society},
  doi = {10.1103/PhysRevB.88.245137},
  url = {https://link.aps.org/doi/10.1103/PhysRevB.88.245137}
}

@article{mao2023sampling,
  title={Sampling reduced density matrix to extract fine levels of entanglement spectrum and restore entanglement Hamiltonian},
  author={Mao, Bin-Bin and Ding, Yi-Ming and Wang, Zhe and Hu, Shijie and Yan, Zheng},
  journal={Nature Communications},
  volume={16},
  number={1},
  pages={2880},
  year={2025},
  publisher={Nature Publishing Group UK London}
}

@article{dalmonte2022entanglement,
  title={Entanglement Hamiltonians: from field theory to lattice models and experiments},
  author={Dalmonte, Marcello and Eisler, Viktor and Falconi, Marco and Vermersch, Beno{\^\i}t},
  journal={Annalen der Physik},
  volume={534},
  number={11},
  pages={2200064},
  year={2022},
  publisher={Wiley Online Library}
}

@article{Li2024Relevant,
  title = {Relevant long-range interaction of the entanglement Hamiltonian emerges from a short-range gapped system},
  author = {Li, Chuhao and Huang, Rui-Zhen and Ding, Yi-Ming and Meng, Zi Yang and Wang, Yan-Cheng and Yan, Zheng},
  journal = {Phys. Rev. B},
  volume = {109},
  issue = {19},
  pages = {195169},
  numpages = {8},
  year = {2024},
  month = {May},
  publisher = {American Physical Society},
  doi = {10.1103/PhysRevB.109.195169},
  url = {https://link.aps.org/doi/10.1103/PhysRevB.109.195169}
}

@article{song2023different,
  title={Different temperature dependence for the edge and bulk of the entanglement Hamiltonian},
  author={Song, Menghan and Zhao, Jiarui and Yan, Zheng and Meng, Zi Yang},
  journal={Physical Review B},
  volume={108},
  number={7},
  pages={075114},
  year={2023},
  publisher={APS}
}

@article{Mermin1966Absence,
  title = {Absence of Ferromagnetism or Antiferromagnetism in One- or Two-Dimensional Isotropic Heisenberg Models},
  author = {Mermin, N. D. and Wagner, H.},
  journal = {Phys. Rev. Lett.},
  volume = {17},
  issue = {22},
  pages = {1133--1136},
  numpages = {0},
  year = {1966},
  month = {Nov},
  publisher = {American Physical Society},
  doi = {10.1103/PhysRevLett.17.1133},
  url = {https://link.aps.org/doi/10.1103/PhysRevLett.17.1133}
}

@article{mermin1967absence,
  title={Absence of ordering in certain classical systems},
  author={Mermin, N David},
  journal={Journal of Mathematical Physics},
  volume={8},
  number={5},
  pages={1061--1064},
  year={1967},
  publisher={American Institute of Physics}
}

@misc{diessel2022generalized,
      title={Generalized Higgs mechanism in long-range interacting quantum systems}, 
      author={Oriana K. Diessel and Sebastian Diehl and Nicolò Defenu and Achim Rosch and Alessio Chiocchetta},
      year={2022},
      eprint={2208.10487},
      archivePrefix={arXiv},
      primaryClass={cond-mat.quant-gas}
}

@article{PhysRevA.103.043321,
  title = {Intercomponent entanglement entropy and spectrum in binary Bose-Einstein condensates},
  author = {Yoshino, Takumi and Furukawa, Shunsuke and Ueda, Masahito},
  journal = {Phys. Rev. A},
  volume = {103},
  issue = {4},
  pages = {043321},
  numpages = {14},
  year = {2021},
  month = {Apr},
  publisher = {American Physical Society},
  doi = {10.1103/PhysRevA.103.043321},
  url = {https://link.aps.org/doi/10.1103/PhysRevA.103.043321}
}

@article{Yusuf2004spin,
  title = {Spin waves in antiferromagnetic spin chains with long-range interactions},
  author = {Yusuf, Eddy and Joshi, Anuvrat and Yang, Kun},
  journal = {Phys. Rev. B},
  volume = {69},
  issue = {14},
  pages = {144412},
  numpages = {7},
  year = {2004},
  month = {Apr},
  publisher = {American Physical Society},
  doi = {10.1103/PhysRevB.69.144412},
  url = {https://link.aps.org/doi/10.1103/PhysRevB.69.144412}
}

@Article{zyan2021entanglement,
author={Yan, Zheng
and Meng, Zi Yang},
title={Unlocking the general relationship between energy and entanglement spectra via the wormhole effect},
journal={Nature Communications},
year={2023},
month={Apr},
day={24},
volume={14},
number={1},
pages={2360},
issn={2041-1723},
doi={10.1038/s41467-023-37756-7},
url={https://doi.org/10.1038/s41467-023-37756-7}
}

@article{yan2021topological,
  title={Topological phase transition and single/multi anyon dynamics of Z 2 spin liquid},
  author={Yan, Zheng and Wang, Yan-Cheng and Ma, Nvsen and Qi, Yang and Meng, Zi Yang},
  journal={npj Quantum Materials},
  volume={6},
  pages={39},
  year={2021},
  publisher={Nature Publishing Group}
}

@article{Xu2019,
  title = {Spin excitation spectra of the two-dimensional $S=\frac{1}{2}$ Heisenberg model with a checkerboard structure},
  author = {Xu, Yining and Xiong, Zijian and Wu, Han-Qing and Yao, Dao-Xin},
  journal = {Phys. Rev. B},
  volume = {99},
  issue = {8},
  pages = {085112},
  numpages = {13},
  year = {2019},
  month = {Feb},
  publisher = {American Physical Society},
  doi = {10.1103/PhysRevB.99.085112},
  url = {https://link.aps.org/doi/10.1103/PhysRevB.99.085112}
}

@article{lieb1961two,
  title={Two soluble models of an antiferromagnetic chain},
  author={Lieb, Elliott and Schultz, Theodore and Mattis, Daniel},
  journal={Annals of Physics},
  volume={16},
  number={3},
  pages={407--466},
  year={1961},
  publisher={Elsevier}
}

@article{oshikawa2000commensurability,
  title={Commensurability, excitation gap, and topology in quantum many-particle systems on a periodic lattice},
  author={Oshikawa, Masaki},
  journal={Physical Review Letters},
  volume={84},
  number={7},
  pages={1535},
  year={2000},
  publisher={APS}
}

@article{hastings2004lieb,
  title={Lieb-Schultz-Mattis in higher dimensions},
  author={Hastings, Matthew B},
  journal={Physical Review B},
  volume={69},
  number={10},
  pages={104431},
  year={2004},
  publisher={APS}
}

@article{ma2018dynamical,
  title = {Dynamical signature of fractionalization at a deconfined quantum critical point},
  author = {Ma, Nvsen and Sun, Guang-Yu and You, Yi-Zhuang and Xu, Cenke and Vishwanath, Ashvin and Sandvik, Anders W. and Meng, Zi Yang},
  journal = {Phys. Rev. B},
  volume = {98},
  issue = {17},
  pages = {174421},
  numpages = {12},
  year = {2018},
  month = {Nov},
  publisher = {American Physical Society},
  doi = {10.1103/PhysRevB.98.174421},
  url = {https://link.aps.org/doi/10.1103/PhysRevB.98.174421}
}

@article{Zhou2021amplitude,
  title = {Amplitude Mode in Quantum Magnets via Dimensional Crossover},
  author = {Zhou, Chengkang and Yan, Zheng and Wu, Han-Qing and Sun, Kai and Starykh, Oleg A. and Meng, Zi Yang},
  journal = {Phys. Rev. Lett.},
  volume = {126},
  issue = {22},
  pages = {227201},
  numpages = {6},
  year = {2021},
  month = {Jun},
  publisher = {American Physical Society},
  doi = {10.1103/PhysRevLett.126.227201},
  url = {https://link.aps.org/doi/10.1103/PhysRevLett.126.227201}
}

@article{Poilblanc2010entanglement,
  title = {Entanglement Spectra of Quantum Heisenberg Ladders},
  author = {Poilblanc, Didier},
  journal = {Phys. Rev. Lett.},
  volume = {105},
  issue = {7},
  pages = {077202},
  numpages = {4},
  year = {2010},
  month = {Aug},
  publisher = {American Physical Society},
  doi = {10.1103/PhysRevLett.105.077202},
  url = {https://link.aps.org/doi/10.1103/PhysRevLett.105.077202}
}

@article{Li2008entangle,
  title = {Entanglement Spectrum as a Generalization of Entanglement Entropy: Identification of Topological Order in Non-Abelian Fractional Quantum Hall Effect States},
  author = {Li, Hui and Haldane, F. D. M.},
  journal = {Phys. Rev. Lett.},
  volume = {101},
  issue = {1},
  pages = {010504},
  numpages = {4},
  year = {2008},
  month = {Jul},
  publisher = {American Physical Society},
  doi = {10.1103/PhysRevLett.101.010504},
  url = {https://link.aps.org/doi/10.1103/PhysRevLett.101.010504}
}

@article{Fradkin2006entangle,
  title = {Entanglement Entropy of 2D Conformal Quantum Critical Points: Hearing the Shape of a Quantum Drum},
  author = {Fradkin, Eduardo and Moore, Joel E.},
  journal = {Phys. Rev. Lett.},
  volume = {97},
  issue = {5},
  pages = {050404},
  numpages = {4},
  year = {2006},
  month = {Aug},
  publisher = {American Physical Society},
  doi = {10.1103/PhysRevLett.97.050404},
  url = {https://link.aps.org/doi/10.1103/PhysRevLett.97.050404}
}

@article{guo2021entanglement,
  title={Entanglement spectrum of geometric states},
  author={Guo, Wu-zhong},
  journal={Journal of High Energy Physics},
  volume={2021},
  number={2},
  pages={1--33},
  year={2021},
  publisher={Springer},
  url={https://link.springer.com/article/10.1007/JHEP02(2021)085}
}

@article{GYSun2018,
	title = {Dynamical Signature of Symmetry Fractionalization in Frustrated Magnets},
	author = {Sun, Guang-Yu and Wang, Yan-Cheng and Fang, Chen and Qi, Yang and Cheng, Meng and Meng, Zi Yang},
	journal = {Phys. Rev. Lett.},
	volume = {121},
	issue = {7},
	pages = {077201},
	numpages = {6},
	year = {2018},
	month = {Aug},
	publisher = {American Physical Society},
	doi = {10.1103/PhysRevLett.121.077201},
	url = {https://link.aps.org/doi/10.1103/PhysRevLett.121.077201}
}

@article{zhu2019reconstructing,
  title = {Reconstructing entanglement Hamiltonian via entanglement eigenstates},
  author = {Zhu, W. and Huang, Zhoushen and He, Yin-Chen},
  journal = {Phys. Rev. B},
  volume = {99},
  issue = {23},
  pages = {235109},
  numpages = {12},
  year = {2019},
  month = {Jun},
  publisher = {American Physical Society},
  doi = {10.1103/PhysRevB.99.235109},
  url = {https://link.aps.org/doi/10.1103/PhysRevB.99.235109}
}

@article{verresen2021gapless,
  title = {Gapless Topological Phases and Symmetry-Enriched Quantum Criticality},
  author = {Verresen, Ruben and Thorngren, Ryan and Jones, Nick G. and Pollmann, Frank},
  journal = {Phys. Rev. X},
  volume = {11},
  issue = {4},
  pages = {041059},
  numpages = {41},
  year = {2021},
  month = {Dec},
  publisher = {American Physical Society},
  doi = {10.1103/PhysRevX.11.041059},
  url = {https://link.aps.org/doi/10.1103/PhysRevX.11.041059}
}

@article{AKLT,
  title = {Rigorous results on valence-bond ground states in antiferromagnets},
  author = {Affleck, Ian and Kennedy, Tom and Lieb, Elliott H. and Tasaki, Hal},
  journal = {Phys. Rev. Lett.},
  volume = {59},
  issue = {7},
  pages = {799--802},
  numpages = {0},
  year = {1987},
  month = {Aug},
  publisher = {American Physical Society},
  doi = {10.1103/PhysRevLett.59.799},
  url = {https://link.aps.org/doi/10.1103/PhysRevLett.59.799}
}

@article{Laflorencie2005Critical,
doi = {10.1088/1742-5468/2005/12/P12001},
url = {https://dx.doi.org/10.1088/1742-5468/2005/12/P12001},
year = {2005},
month = {dec},
publisher = {},
volume = {2005},
number = {12},
pages = {P12001},
author = {Laflorencie, Nicolas and Affleck, Ian and Berciu, Mona},
title = {Critical phenomena and quantum phase transition in long range Heisenberg antiferromagnetic
chains},
journal = {Journal of Statistical Mechanics: Theory and Experiment},

}

@article{zhao2025Unconventional,
  title = {Unconventional Scalings of Quantum Entropies in Long-Range Heisenberg Chains},
  author = {Zhao, Jiarui and Laflorencie, Nicolas and Meng, Zi Yang},
  journal = {Phys. Rev. Lett.},
  volume = {134},
  issue = {1},
  pages = {016707},
  numpages = {6},
  year = {2025},
  month = {Jan},
  publisher = {American Physical Society},
  doi = {10.1103/PhysRevLett.134.016707},
  url = {https://link.aps.org/doi/10.1103/PhysRevLett.134.016707}
}

@article{Song2023Dynamical,
   title={Dynamical properties of quantum many-body systems with long-range interactions},
   volume={5},
   issue = {3},
   pages = {033046},
   ISSN={2643-1564},
   url={http://dx.doi.org/10.1103/PhysRevResearch.5.033046},
   number={3},
   journal={Physical Review Research},
   publisher={American Physical Society (APS)},
   author={Song, Menghan and Zhao, Jiarui and Zhou, Chengkang and Meng, Zi Yang},
   year={2023},
   month={July} }

@misc{yang2024dynamics,
      title={Dynamic structure factor of a spin-1/2 Heisenberg chain with long-range interactions}, 
      author={Sibin Yang and Gabe Schumm and Anders W. Sandvik},
      year={2024},
      eprint={2412.15168},
      archivePrefix={arXiv},
      primaryClass={cond-mat.str-el},
      url={https://arxiv.org/abs/2412.15168}, 
}

@article{Lieb1962Ordering,
    author = {Lieb, Elliott and Mattis, Daniel},
    title = {Ordering Energy Levels of Interacting Spin Systems},
    journal = {Journal of Mathematical Physics},
    volume = {3},
    number = {4},
    pages = {749-751},
    year = {1962},
    month = {07},
    
    issn = {0022-2488},
    doi = {10.1063/1.1724276},
    url = {https://doi.org/10.1063/1.1724276},
    eprint = {https://pubs.aip.org/aip/jmp/article-pdf/3/4/749/19167430/749\_1\_online.pdf},
}

@article{Yu2024Universal,
  title = {Universal Entanglement Spectrum in One-Dimensional Gapless Symmetry Protected Topological States},
  author = {Yu, Xue-Jia and Yang, Sheng and Lin, Hai-Qing and Jian, Shao-Kai},
  journal = {Phys. Rev. Lett.},
  volume = {133},
  issue = {2},
  pages = {026601},
  numpages = {9},
  year = {2024},
  month = {Jul},
  publisher = {American Physical Society},
  doi = {10.1103/PhysRevLett.133.026601},
  url = {https://link.aps.org/doi/10.1103/PhysRevLett.133.026601}
}

@article{Parker2018Topological,
  title = {Topological Luttinger liquids from decorated domain walls},
  author = {Parker, Daniel E. and Scaffidi, Thomas and Vasseur, Romain},
  journal = {Phys. Rev. B},
  volume = {97},
  issue = {16},
  pages = {165114},
  numpages = {10},
  year = {2018},
  month = {Apr},
  publisher = {American Physical Society},
  doi = {10.1103/PhysRevB.97.165114},
  url = {https://link.aps.org/doi/10.1103/PhysRevB.97.165114}
}

@article{Scaffidi2017Gapless,
  title = {Gapless Symmetry-Protected Topological Order},
  author = {Scaffidi, Thomas and Parker, Daniel E. and Vasseur, Romain},
  journal = {Phys. Rev. X},
  volume = {7},
  issue = {4},
  pages = {041048},
  numpages = {16},
  year = {2017},
  month = {Nov},
  publisher = {American Physical Society},
  doi = {10.1103/PhysRevX.7.041048},
  url = {https://link.aps.org/doi/10.1103/PhysRevX.7.041048}
}

@misc{liu2026zenodo,
  author       = {Liu, Zenan and
                  Wang, Zhe and
                  Yao, Dao-Xin and
                  Yan, Zheng},
  title        = {Data for "Worldline deconfinement and emergent
                   long-range interaction in entanglement Hamiltonian
                   and entanglement spectrum"
                  },
  month        = mar,
  year         = 2026,
  publisher    = {Zenodo},
  doi          = {10.5281/zenodo.19084350},
  url          = {https://doi.org/10.5281/zenodo.19084350},
}

\end{document}